\documentclass[a4paper, 12pt, 3p]{elsarticle}

\usepackage{multicol}
\usepackage{color}
\usepackage{xspace}
\usepackage{pdfwidgets}
\usepackage{enumerate}
\usepackage[export]{adjustbox}
\usepackage{graphicx}
\usepackage{tikz}
\usepackage{subfigure}
\usepackage{amsmath}
\usepackage{multicol}
\usepackage{gensymb}
\usepackage{dcolumn}
\usepackage{bm}
\usepackage{hyperref}
\hypersetup{colorlinks=true}
\usepackage{multirow}
\usepackage{revsymb}
\usepackage{amssymb}
\usepackage[nodots]{numcompress}
\usepackage{natbib}

\begin{document}

\journal{Physica Medica}

\begin{frontmatter}


\title{Virtual extrapolation technique for retracing line of response of single scattered events in positron emission tomography}


\author{Satyajit Ghosh}
\author{Pragya Das\corref{cor1}}
\ead{pragya@phy.iitb.ac.in}
\address{Department of Physics, Indian Institute of Technology Bombay, Mumbai-400076, India}
\cortext[cor1]{Corresponding author}

\begin{graphicalabstract}

\includegraphics[width=0.85\textwidth, height=0.85\textheight,keepaspectratio] {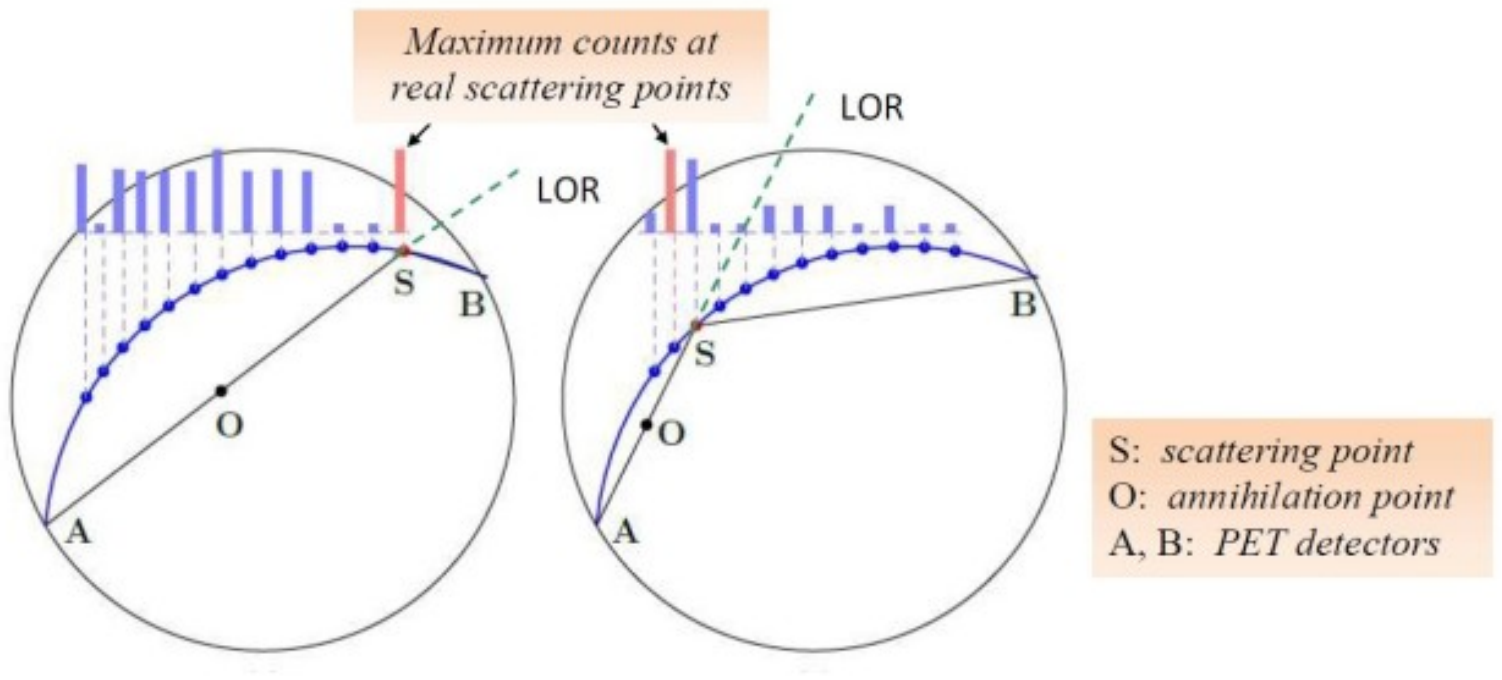}
\end{graphicalabstract}

\begin{highlights}
\item Compton scattering correction in Positron emission tomography
\item Probability density functions for single and two-simultaneous scattering phenomena
\item Monte Carlo simulation for coincidence events in list mode 
\item New Virtual extrapolation technique for converting two-parameter list mode data to one-parameter data set
\item Recognizing real scattering points using a Collective difference property

\end{highlights}
\begin{abstract}

\noindent {\bf Purpose:} 
The scattering phenomenon creates degrading effects in positron emission tomography (PET) and the corresponding events are rejected conventionally. We have proposed a mathematical model to retrace the original line of response of the single-scattered coincident events with the aim to incorporate such events in PET. \\
\noindent {\bf Methods:}
We have devised a new Virtual extrapolation technique based on the concept of the probability density functions. Through which we transformed the original two-parameter list mode data for the coincident photon pairs to a one-parameter data set. The procedure of random sampling and sampling distribution -- by utilizing some unique properties like a collective difference and the length compensation -- was employed in the data analysis. We studied the effect of finite timing and energy resolution of detectors on the performance of the proposed model.
\\
\noindent {\bf Results:}
We determined the frequency of occurrence of the data values corresponding to real as well as fictitious scattering points -- placed on the arc of a circle drawn with the constant scattering angle -- to observe the highest counts. As expected, we found the highest counts at the location of real scattering points. The model was evaluated for a uniform attenuating phantom medium. The results were found impressive for the case of ideal time and energy information, less so for the cases of finite resolutions.
 \\
\noindent {\bf Conclusions:}
Our work outlines a completely new approach; though, a lot more sophistication is required for its practical utility.  Nonetheless, the technique seems promising in developing a potential approach to improve the sensitivity of PET imaging.

\end{abstract}


\end{frontmatter}

\section{\large Introduction}
\label{sec:level1}
The scattering phenomenon in positron emission tomography (PET) produces undesirable effects in the reconstructed image. It degrades image contrast, signal-to-noise ratio (SNR), and quantitative accuracy of the image. The fact that the line of response (LOR) for scattered coincident events never contains annihilation point, those events are never used directly in image reconstruction. In conventional PET, the scattered events are first estimated and carefully rejected from the data~\cite{zaidi2007scatter} for the unscattered part to be successfully utilized in image reconstruction. Various experimental and simulation-based techniques have been developed over the years for estimating and subtracting scattered events, \textit{e.g.}, use of multiple energy windows in conjunction with the photo-peak window \cite{shao1994triple,grootoonk1996correction}, Monte Carlo \cite{levin1995monte}, and single scatter simulation approach \cite{watson1996single}.\par

In a typical PET scan, Compton scattering contributes maximum to the total scattering interactions ($>$ 99.7\% for water medium) \cite{zaidi2007strategies}. Out of that, more than 80\% are due to single scattering \cite{zaidi2004scatter,bailey2005positron}. Scatter fraction depends upon various factors -- position and width of the energy window, patient size, two or three-dimensional acquisition mode, and the use of lead or tungsten septa. In a typical 2D scan, the scatter fraction can be 10-20\% of the total data. Whereas in 3D, it can go up to 30-60\% \cite{zaidi2007scatter}. Since the maximum sensitivity attainable in a typical emission tomographic scan is less than 5\% of the total emitted photons from the patient body \cite{reader20144d}, rejecting scattered portion from the sparse total data amplifies noise in the reconstructed image.\par      

The use of Compton scattered events in medical imaging and non-destructive testing (NDT) has been popular for a long time. This subfield of tomographic reconstruction is called `Compton scattered tomography' (CST) \cite{truong2012recent}. The mathematical foundation of CST was initiated by Allan Cormack \cite{cormack1981radon,cormack1982radon} who studied Radon transform \cite{deans2007radon} on a family of curves on a 2D plane. Besides, he established the usefulness of conventional straight-line Radon transform in medical imaging. Norton~\cite{norton1994compton}, Troung \textit{et al.}~\cite{truong2007mathematical}, and Nguyen \textit{et al.}~\cite{nguyen2009novel} developed various generalized versions of Radon transform. They evaluated the usefulness of these newly developed transformations in gamma camera systems. However, in PET, due to highly complex data acquisition, the usefulness of newly developed transformations is still under initial investigation. On the other hand, recently, various researchers are interested in the feasibility study of using scattered events through a modified iterative reconstruction technique. A simulation-based feasibility study of using both single scattered and unscattered data in image reconstruction was proposed by Conti \textit{et al.}~\cite{conti2012reconstruction} for the time-of-flight PET (TOF-PET) systems. Sun \textit{et al.}~\cite{sun2013evaluation} developed a `generalized scatter maximum-likelihood expectation-maximization' (GS-MLEM) algorithm for the non-TOF-PET systems, which includes both unscattered and scattered events in image reconstruction. Similar to GS-MLEM, another generalized iterative algorithm was proposed by Hemmati \textit{et al.}~\cite{hemmati2017compton} for the TOF-PET systems.\par

List mode data acquisition in emission tomography opens up a way to use energies and time-of-flight (TOF) information (if available) of individual events. Many researchers have proposed efficient techniques for estimating and rejecting scattered events using list mode data \cite{hemmati2017compton,popescu2006pet,guerin2010novel}. With the advent of detectors possessing good energy and timing resolutions, the problem of retracing original LOR for the scattered event has become tractable. Instead of rejecting scattered events, adding them to image reconstruction can improve SNR. Also, one can retain good sensitivity and image contrast with less noise, even with the reduced dose of radioactive tracers. Besides, the prospect of reduced statistical noise in a data scan can improve the quality of tracer kinetic modeling~\cite{bailey2005positron,karakatsanis2013dynamic}.\par

In the present work, we have attempted to retrace the LOR of single scattered coincident events outside the image reconstruction process. Our approach relies on an entirely different concept as compared to earlier works. At first, we derived a distribution function (W) for the `double scattering' of the coincident photon pairs occurring in real as well as in a virtual uniform attenuating medium of an infinite extent.  The meaning of `double scattering' has to be understood in context; that it does not correspond to two simultaneous real scatterings. The implication is that we have been consistent with our aim of studying `single scattered' events mentioned in the title of the present work. We employed two unique features -- a data transformation based on the `Virtual extrapolation technique', and the `collective difference property' of the transformed data. The immediate consequence of data transformation was the conversion of every event of double coincidence list mode data (two-parameter) to singles data (one-parameter), simplifying the data structure. Moreover, the transformation process was such that among all possible scattering points on the loci of two circular arcs~\cite{conti2012reconstruction}, only data values corresponding to the original (real) scattering points obeyed the distribution function W. Using the concepts of random sampling and sampling distribution for analyzing the transformed data, the LORs of the single scattered coincident events were retraced by recognizing the original scattering points on the circular loci. Our technique can be applied to both TOF and non-TOF-PET systems. We have also demonstrated the performance of our model for the case of finite time and energy resolution of detectors. The proposed technique has been evaluated for a phantom of uniform attenuation. But, our model can be easily extended to a non-uniform phantom, as the attenuation map is generally known apriori using a CT or MRI scanner~\cite{kinahan2003,judenhofer2008simultaneous}.
\par

We have succeeded in identifying the original scattering point among all possible points within a reasonable error (Sec.~\ref{sec:level4}) for a uniform phantom medium. As the work was in an early stage of development and currently cannot handle total PET scan data altogether, we have avoided showing any enhancement in the quality of the reconstructed image. Also, the proposed data analysis relied on some empirical understandings (Sec.~\ref{sec:level6}). We, on the other hand, have focused on developing the model and a data analysis algorithm, while investigating thoroughly the ideal, and cases of finite time and energy resolutions. We believe our model could be of great importance for a 3D phantom medium with a relatively higher scatter fraction~\cite{zaidi2007scatter} than for a 2D case, and we can utilize the same double scattering distribution function (W) described in the present work. Our preliminary results, in part, were presented as e-poster in the World Congress on Medical Physics \& Biomedical Engineering, 2018~\cite{satyajit2018prague}.

\section{\large Methods}
\label{sec:level2}

The Virtual extrapolation technique was our new way of transforming PET scan data. In a real PET scan, the phantom medium is of finite size, and that is where the new concept -- the virtual attenuating medium of infinite extent for virtual scattering -- becomes important. The the first few sections outline how we arrived at the proposed technique while maintaining much simplicity for a complex problem.\par 

Our proposed data analysis algorithm focused on using a unique property, named as collective difference property', to achieve the goal of recognizing real scattering points inside the PET detector ring. We first analyzed the data for perfect energy and timing information, and extended the idea to cases of finite time and energy resolutions as well.\par 

\subsection{\bf{Mathematical model}}
\label{sec:level2_1}

We used the basic mathematical tool of probability density function (PDF) for the attenuating path length of photons in a uniform medium and applied initially to single scattering. Next, we developed the concept of coincident double scattering to be utilized later for the scattering of two photons in a virtual or real medium.

\subsubsection{Probability density function for single scattering}
\label{sec:level2_1_1}

We know from the Beer-Lambert law \cite{hsieh2003computed}
\[
N(l)=N_{0} \ e^{-\mu \ l},
\]
where $N_{0}$ is the initial number of photons, $N_{l}$ is the number survived after traveling a distance $l$, and $\mu$ is the attenuation coefficient of a uniform attenuating medium at 511 keV photon energy. Differentiating both sides we get,
\[
dN(l)=-N_{0} \ \mu \ e^{-\mu \ l} \ dl.
\]
So,
\[
\frac{-\frac{dN}{N_{0}}}{dl}=\mu \ e^{-\mu \ l}.
\]

Above expression directly gives the probability density function for distance ($l$) between a source point and the first Compton scattering point of photon,
\begin{eqnarray}\label{equ1}
W(l)=\mu \ e^{-\mu \ l}.
\end{eqnarray}

Here we have replaced `attenuating length' with `first Compton scattering length' because it dominates in PET~\cite{zaidi2007strategies} in comparison to photoelectric absorption and coherent scattering.

\subsubsection{Probability density function for coincident double scattering}
\label{sec:level2_1_2}

Having found the probability density function (PDF) for the single scattering~(Eq.~\ref{equ1}), we extended the idea to double scattering of coincident photons. The finite positron travel path and non-collinearity of $e^{-}$- $e^{+}$ annihilation were discarded as they affect negligibly on the spatial resolution of the reconstructed images in a typical human body PET scanner \cite{sanchez2004positron,cherry2012physics}. Figure \ref{fig1} presents a positron emitting point source placed at $O$ inside a uniform attenuating medium of `infinite extent'. Consider $S_{1}$ and $S_{2}$ as two scattering points for a coincident event, and let $\overline{OS_{1}}$ be $l$ and $\overline{S_{1}OS_{2}}$ be $z$, the corresponding PDF is
 \begin{eqnarray}\label{equ2}
W(z)=\int_{0}^{z}\mu \ e^{-\mu(z-l)} \ \mu \ e^{-\mu \ l}dl = {\mu}^{2} \ z \ e^{-\mu z}
\end{eqnarray}  
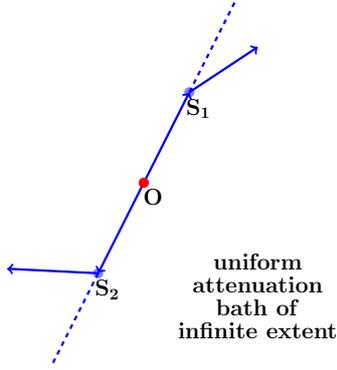
\begin{figure}[!htb]
    \centering
        \scalebox{0.6}{\begin{tikzpicture}
\coordinate[label={[below]\textbf{\large{$\mathbf{O}$}}}] (a) at (0.2,-0.0);
\coordinate[label={[below]\textbf{\large{$\mathbf{S_{1}}$}}}] (b) at (1.2,2);
\coordinate[label={[below]\textbf{\large{$\mathbf{S_{2}}$}}}] (c) at (-0.8,-2);
\filldraw[color=red!50,fill=red!50,very thin](0,0) circle (0.1cm);
\filldraw[color=blue!50,fill=blue!50](1,2) circle (0.1cm);
\filldraw[color=blue!50,fill=blue!50](-1,-2) circle (0.1cm);
\draw[blue!100,line width=0.5mm,solid,->](0,0) -- (1,2);
\draw[blue!100,line width=0.5mm,solid,->](0,0) -- (-1,-2);
\draw[blue!100,line width=0.5mm,dashed](1,2) -- (2,4);
\draw[blue!100,line width=0.5mm,dashed](-1,-2) -- (-2,-4);
\draw[blue!100,line width=0.5mm,solid,->](1,2) -- (2.5,3);
\draw[blue!100,line width=0.5mm,solid,->](-1,-2) -- (-3,-1.9);
\filldraw[color=red!100,fill=red!100,very thin](0,0) circle (0.1cm);
\node[align=center] at (2.5,-2.5) {\large{\textbf{uniform}}\\ \large{\textbf{attenuation}}\\ \large{\textbf{bath of}}\\ \large{\textbf{infinite extent}}};
\end{tikzpicture}} 
      \caption{Double scattering of a coincident photon pair inside a uniform attenuating medium of infinite extent.}
      \label{fig1}
\end{figure}

\begin{figure}[!htb]  
\centering
\includegraphics[width=0.53\textwidth, height=0.53\textheight,keepaspectratio]{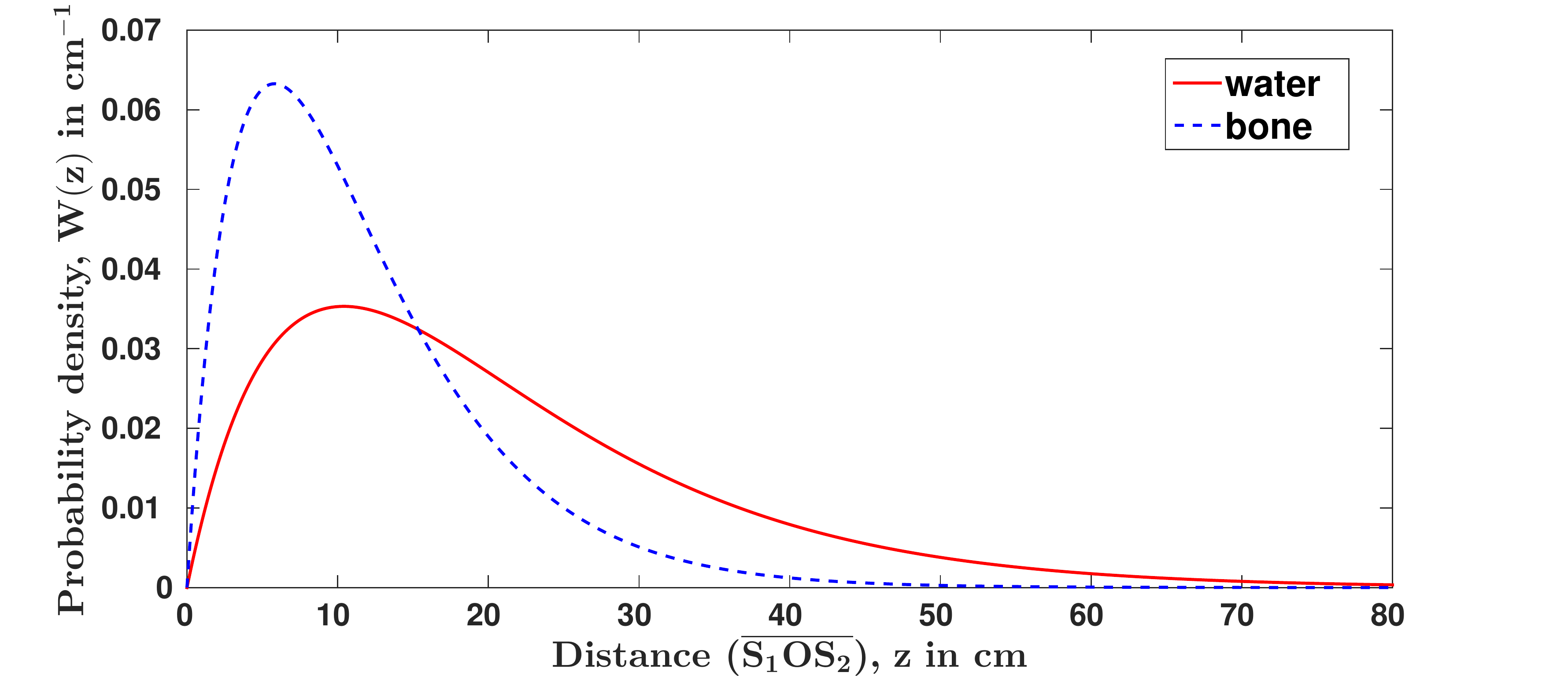}
\caption{Probability density functions of the double-scattering distances ($z$) for the coincident photon pair of energy 511 keV, inside water and bone mediums.}
\label{fig2}
\end{figure}

The expression~(\ref{equ2}) relies on the fact that the PDF of sum of two independent random variables -- true for the attenuating path lengths of two coincident photons -- is the convolution of PDFs of each of the two random variables \cite{feller2008introduction}. Figure~\ref{fig2} shows the plots of $W(z)$ for two attenuating mediums -- water and bone. At 511 keV, the attenuation coefficients of water and bone are 0.096 $cm^{-1}$ and 0.172 $cm^{-1}$, respectively \cite{hubbell1995tables}. We applied the very idea of double coincidence scattering~(Eq.~\ref{equ2}) for the `virtual scattering' in a `virtual attenuating medium' and `real scattering' in `real medium', described in the next section.\par
Since the attenuating medium is uniform, the same distribution function $W(z)$~(Eq.~\ref{equ2}) applies irrespective of the position of source point, also same for a distributed radioactive source. Moreover, the derived distribution function remains the same in three dimensional attenuating medium also. Hence, from a theoretical perspective, the proposed model is equally applicable for a 3D-PET system.\par

\subsection{\bf{Virtual Extrapolation}}
\label{sec:level2_2}

The challenge for us in developing the model was to find a connectivity between the real PET data and that of PDF $W(z)$ (Eq.~\ref{equ2}). We have achieved this through our newly devised Virtual extrapolation technique which transforms all PET coincident events one by one into the double-scattering events of the same distribution function W(z). \par

\subsubsection{Unscattered and single scattered events}
\label{sec:level2_2_1}

\begin{figure*}[htp]
  \centering
  \subfigure[]{\scalebox{0.5}{\begin{tikzpicture}
\path[fill=gray!35,even odd rule]
  (-6,-6.8) rectangle (7.2,3.0)
  (1.5,-1.5) circle (3.5cm);
\draw[line width=0.4mm,dashed,blue!100](1.5,-1.5) circle (3.47cm);
\filldraw[color=gray!90,fill=gray!10,very thick](0,-3) rectangle (3,0);
\filldraw[color=gray!90,fill=gray!30,very thick](1.5,-1.5) ellipse (1.2cm and 0.9 cm);
\filldraw[color=gray!90,fill=black!30,very thick](1.7,-1.9) circle (0.4cm);
\draw[line width=0.5mm,green!100,->](1,-1) -- (4.1047,0.7927);
\draw[line width=0.5mm,green!100,->](1,-1) -- (-1.7878,-2.6097);
\filldraw[color=blue!100,fill=blue!100,very thick](5,1.3096) circle (0.1cm);
\filldraw[color=blue!100,fill=blue!100,very thick](-3.5,-3.5983) circle (0.1cm);
\draw[line width=0.5mm,green!100,dashed,->](4.1047,0.7927) -- (5,1.3096);
\draw[line width=0.5mm,green!100,dashed,->](-1.7878,-2.6097) -- (-3.5,-3.5983);
\draw[line width=0.5mm,green!100,dashed,->](5,1.3096) -- (7,1);
\draw[line width=0.5mm,green!100,dashed,->](-3.5,-3.5983) -- (-5.5,-3.5);
\coordinate[label={[below]\textbf{\large A}}] (a) at (-1.9,-2.8);
\coordinate[label={[below]\textbf{\large B}}] (b) at (4.1047,0.65);
\coordinate[label={[below]{\large $\mathbf{A^{'}}$}}] (a) at (-0.4,-0.9);
\coordinate[label={[below]{\large $\mathbf{B^{'}}$}}] (b) at (2.6,0.8);
\coordinate[label={[below]\textbf{\large C}}] (c) at (5,2.0);
\coordinate[label={[below]\textbf{\large D}}] (d) at (-3.5,-3.7);
\coordinate[label={[above]\textbf{\large O}}] (e) at (1,-1.75);
\filldraw[color=red!100,fill=red!100,very thick](1,-1) circle (0.1cm);
\node[align=center] at (5.1,-5.3) { \bf{\large uniform}\\ \bf{\large attenuation}\\ \bf{\large bath of}\\ \bf{\large infinite extent}};
\end{tikzpicture}}}\quad
  \subfigure[]{\scalebox{0.23}{\input{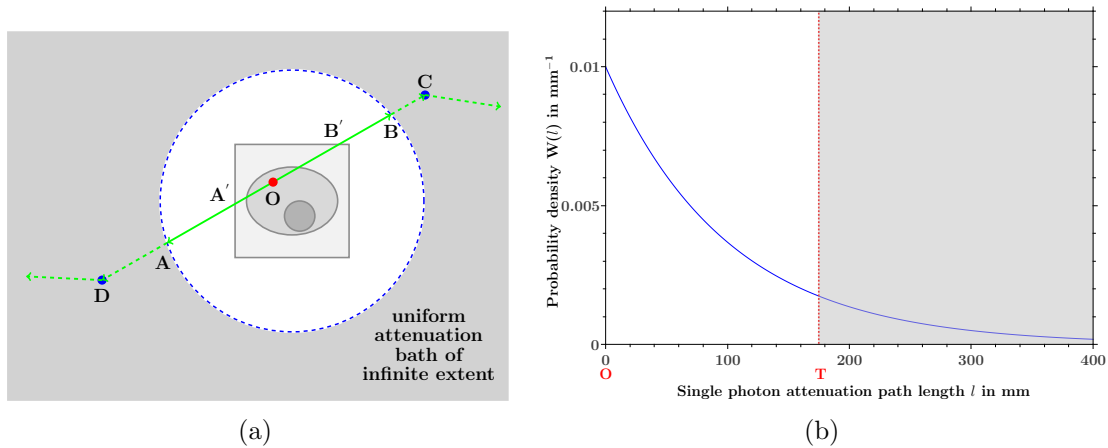}}}
  \caption{(a) A typical PET detector ring and a square radioactive intensity distribution depicting the variation in intensity with different shades of gray, also showing an annihilation point $O$.  For simplicity in demonstration, a uniform attenuating medium is considered filling inside the whole space of the PET ring. Points $A$, $B$ are the detection points of two photons of an unscattered coincident event, and $C$, $D$ are the corresponding assigned virtual scattering points, described in Sec.~\ref{sec:level2_2_1}. Also, points $A^{'}$ and $B^{'}$ are the cross-points of the line of response on the activity distribution boundary (Sec.~\ref{sec:level3_1}). (b) A plot of the probability density function $W(l)$ (Eq.~\ref{equ1}) for the single photon attenuating path length $l$, with $T$ marked at the detection point of the photon, and the shaded portion representing a region of virtual scattering points.}
  \label{fig3}
\end{figure*}

At first, we describe the transformation procedure for PET data corresponding to unscattered coincidences. For such photons, without any real scattering, we identified the virtual (or fictitious) scattering points outside the PET detector ring. The virtual medium outside the PET ring was assumed to be of the same attenuation coefficient as inside. Figure~\ref{fig3}(a) shows an annihilation point $O$ and two coincident photons detected at points $A$ and $B$ without any scattering inside the phantom medium. Square shape in the figure denotes the radioactive intensity distribution, and for simplicity in demonstration we have considered a uniform attenuating medium filled inside the whole space of PET detector ring. The time-of-flight (TOF-PET) information fixes the position of the annihilation point $O$ uniquely, deciding the path lengths $OA$ and $OB$. For the non-TOF-PET, the position of $O$ is arbitrary but can be fixed, lying anywhere on the line of response $AB$. Interestingly, the time-of-flight information is redundant in our approach and can be qualitatively understood by the fact that we are dealing with coincident events. If a wrong placement of annihilation point ($O$) reduces the distance $\overline{OA}$, accordingly, the distance $\overline{OB}$ is increased by the same amount and the net effect is zero in Eq.~\ref{equ2}. We termed it ``concept of length compensation'', described in detail in the \ref{appen1}.\par

To identify the virtual scattering points, we generated random values of $l$ -- corresponding to single photon scattering lengths for PDF of Eq.~\ref{equ1} -- by inverse cumulative density function (CDF) method \cite{dunn2011exploring}. Figure~\ref{fig3}(b) shows a plot of PDF of Eq.~\ref{equ1}, marked at $T$ on the x-axis such that the lengths $OT$ and $OA$~(Fig.~\ref{fig3}(a)) are the same. Consequently, all values of $l$ greater than $OT$ (shaded region) were possible for virtual scattering points.  We, therefore, chose one such $l$ value randomly to assign the virtual scattering point $D$ in Fig.~\ref{fig3}(a). Similarly, corresponding to the detection point $B$, the virtual scattering point $C$ was assigned. We proposed the distance ($z$) between $C$ and $D$ to be the transformed data value for the PDF ($W(z)$) of double coincidence scattering (Eq.~\ref{equ2}). We verified our proposition later in Sec.~\ref{sec:level3_2}. We named it `Virtual extrapolation technique' as it extrapolates fictitious scattering points in the virtual medium.  By the process of such transformation, we generated a list-mode data set with one parameter ($z$). \par
 
for the single scattered events, the `real' scattering happened at an unknown location for one photon out of the two, in the coincident pair. Using the Virtual extrapolation technique, we determined the virtual scattering point for the other companion (coincident) unscattered photon following the same procedure as in the previous paragraph.  Having done the Virtual extrapolation for all the unscattered and single scattered events, we got the transformed data set which was used later in identifying the `real' scattering points.\par

To sum up, we identified virtual scattering points for those photons which did not get really Compton scattered inside the phantom. We ignored all sources of error in real PET scan, \textit{e.g.}, random coincidence, error in the estimation of the photon detection points ($A$, $B$, ...). Indeed, due to finite-energy resolution, event-by-event separation of PET data into different parts of the coincidences -- unscattered, single-scattered, and multi-scattered -- cannot be achieved with 100\% accuracy. We considered such errors to be negligible in our present study constituting only the first-stage of developmental work. In this context, recently an innovative way, based on the `polarization-based coincidence event discrimination', of differentiating scattered and unscattered events was proposed~\cite{mcnamara2013positron,mcnamara2014towards,toghyani2016polarisation}.

\subsubsection{multiscattered events}
\label{sec:level2_2_2}

For the multi-scattered coincident events, scattering happened for both the coincident photons once or more, and the information about the locus of scattering points was apparently lost or very hard to obtain. Hence, we avoided performing Virtual extrapolation for such events in the present study. Instead, we obtained the corresponding transformed data directly from the simulated data by calculating the distance between the real scattering points situated inside the PET ring. We justified our way of producing such data as they constituted a very small fraction of the total data, causing minimal error.  Another possible way could be using the first estimate image of the activity distribution produced from the unscattered data.

\subsection{\bf{Monte Carlo simulation of PET data}}
\label{sec:level3_1}

In the step by step procedure of simulating and transforming the PET scan data, we needed to classify all the coincident events into three categories - unscattered events (USE), single scattered events (SSE), and multi-scattered events (MSE). Although our work relied on the simulated PET data, not the real PET scan data, we have adhered to rigorous randomization in three stages using Monte Carlo methods. We believe to have created the data without any bias, equivalent to any real PET scan data. The Virtual extrapolation technique transformed the simulated list mode data to one-parameter data to be used for all further analysis and discussion.\par 

The PET data were simulated through Monte Carlo procedure using C++ codes written by us. Although well established Monte Carlo packages, \textit{e.g.}, GATE~\cite{jan2004gate}, SimSET~\cite{harrison1993preliminary} exist, we used our own codes due to two reasons. Firstly, it was easy enough to write for the physics part of the proposed model. Secondly, we preferred writing the program with all possible checks for our own data analysis algorithm (Sec.~\ref{sec:level4}) as it was in early-stage of development. We, thus, avoided extra complexities and associated errors with ready-made packages.
\par

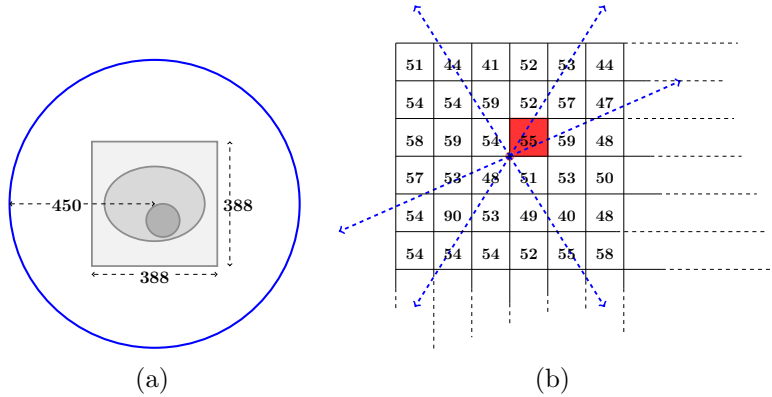
\begin{figure*}[htp]
  \centering
  \subfigure[]{\scalebox{0.55}{\begin{tikzpicture}
\draw[line width=0.5mm,solid,blue!100](1.5,-1.5) circle (3.47cm);
\filldraw[color=gray!90,fill=gray!10,very thick](0,-3) rectangle (3,0);
\draw[line width=0.1mm,dashed,->](1,-3.2) -- (0,-3.2);
\draw[line width=0.1mm,dashed,->](2,-3.2) -- (3,-3.2);
\coordinate[label={[below]\textbf{$\mathbf{388}$}}] (c) at (1.5,-3);
\draw[line width=0.1mm,dashed,->](3.3,-1.3) -- (3.3,0);
\draw[line width=0.1mm,dashed,->](3.3,-1.7) -- (3.3,-3);
\coordinate[label={[below]\textbf{$\mathbf{388}$}}] (d) at (3.5,-1.25);
\filldraw[color=gray!90,fill=gray!30,very thick](1.5,-1.5) ellipse (1.2cm and 0.9 cm);
\filldraw[color=gray!90,fill=black!30,very thick](1.7,-1.9) circle (0.4cm);
\draw[line width=0.1mm,dashed,->](-0.2,-1.5) -- (1.5,-1.5);
\draw[line width=0.1mm,dashed,->](-1,-1.5) -- (-1.97,-1.5);
\coordinate[label={[below]\textbf{$\mathbf{450}$}}] (d) at (-0.60,-1.25);
\end{tikzpicture}}}\quad
  \subfigure[]{\scalebox{0.50}{\begin{tikzpicture}
\filldraw[color=red!80,fill=red!80,very thick](0,-3) rectangle (1,-2);
\filldraw[color=red!50,fill=red!50,very thick](0,-3) circle (0.1cm);
\draw[blue,line width=0.5mm,dashed,->](0,-3) -- (4.5,-1);
\draw[blue,line width=0.5mm,dashed,->](0,-3) -- (-4.5,-5);
\draw[blue,line width=0.5mm,dashed,->](0,-3) -- (2.5,1);
\draw[blue,line width=0.5mm,dashed,->](0,-3) -- (-2.5,-7);
\draw[blue,line width=0.5mm,dashed,->](0,-3) -- (-2.5,1);
\draw[blue,line width=0.5mm,dashed,->](0,-3) -- (2.5,-7);
\draw[black](-3,0) -- (3,0);
\draw[dashed](3,0) -- (6,0);
\draw[black](-3,-1) -- (3.6,-1);
\draw[dashed](3.6,-1) -- (5.6,-1);
\draw[black](-3,-2) -- (2.9,-2);
\draw[dashed](2.9,-2) -- (5.9,-2);
\draw[black](-3,-3) -- (3.7,-3);
\draw[dashed](3.7,-3) -- (6.1,-3);
\draw[black](-3,-4) -- (3.9,-4);
\draw[dashed](3.9,-4) -- (6.5,-4);
\draw[black](-3,-5) -- (2.8,-5);
\draw[dashed](2.8,-5) -- (6.7,-5);
\draw[black](-3,-6) -- (4,-6);
\draw[dashed](4,-6) -- (6.9,-6);
\draw[black](-3,0) -- (-3,-6.5);
\draw[dashed](-3,-6.5) -- (-3,-7.1);
\draw[black](-2,0) -- (-2,-6.1);
\draw[dashed](-2,-6.1) -- (-2,-8.1);
\draw[black](-1,0) -- (-1,-6.5);
\draw[dashed](-1,-6.5) -- (-1,-7.8);
\draw[black](0,0) -- (0,-6.9);
\draw[dashed](0,-6.9) -- (0,-7.5);
\draw[black](1,0) -- (1,-6.6);
\draw[dashed](1,-6.6) -- (1,-7.2);
\draw[black](2,0) -- (2,-6.3);
\draw[dashed](2,-6.3) -- (2,-7.1);
\draw[black](3,0) -- (3,-6.5);
\draw[dashed](3,-6.5) -- (3,-7.7);
\coordinate[label={[below]\textbf{$\mathbf{51}$}}] (a) at (-2.5,-0.3);
\coordinate[label={[below]\textbf{$\mathbf{44}$}}] (a) at (-1.5,-0.3);
\coordinate[label={[below]\textbf{$\mathbf{41}$}}] (a) at (-0.5,-0.3);
\coordinate[label={[below]\textbf{$\mathbf{52}$}}] (a) at (0.5,-0.3);
\coordinate[label={[below]\textbf{$\mathbf{53}$}}] (a) at (1.5,-0.3);
\coordinate[label={[below]\textbf{$\mathbf{44}$}}] (a) at (2.5,-0.3);
\coordinate[label={[below]\textbf{$\mathbf{54}$}}] (a) at (-2.5,-1.3);
\coordinate[label={[below]\textbf{$\mathbf{54}$}}] (a) at (-1.5,-1.3);
\coordinate[label={[below]\textbf{$\mathbf{59}$}}] (a) at (-0.5,-1.3);
\coordinate[label={[below]\textbf{$\mathbf{52}$}}] (a) at (0.5,-1.3);
\coordinate[label={[below]\textbf{$\mathbf{57}$}}] (a) at (1.5,-1.3);
\coordinate[label={[below]\textbf{$\mathbf{47}$}}] (a) at (2.5,-1.3);
\coordinate[label={[below]\textbf{$\mathbf{58}$}}] (a) at (-2.5,-2.3);
\coordinate[label={[below]\textbf{$\mathbf{59}$}}] (a) at (-1.5,-2.3);
\coordinate[label={[below]\textbf{$\mathbf{54}$}}] (a) at (-0.5,-2.3);
\coordinate[label={[below]\textbf{$\mathbf{55}$}}] (a) at (0.5,-2.3);
\coordinate[label={[below]\textbf{$\mathbf{59}$}}] (a) at (1.5,-2.3);
\coordinate[label={[below]\textbf{$\mathbf{48}$}}] (a) at (2.5,-2.3);
\coordinate[label={[below]\textbf{$\mathbf{57}$}}] (a) at (-2.5,-3.3);
\coordinate[label={[below]\textbf{$\mathbf{53}$}}] (a) at (-1.5,-3.3);
\coordinate[label={[below]\textbf{$\mathbf{48}$}}] (a) at (-0.5,-3.3);
\coordinate[label={[below]\textbf{$\mathbf{51}$}}] (a) at (0.5,-3.3);
\coordinate[label={[below]\textbf{$\mathbf{53}$}}] (a) at (1.5,-3.3);
\coordinate[label={[below]\textbf{$\mathbf{50}$}}] (a) at (2.5,-3.3);
\coordinate[label={[below]\textbf{$\mathbf{54}$}}] (a) at (-2.5,-4.3);
\coordinate[label={[below]\textbf{$\mathbf{90}$}}] (a) at (-1.5,-4.3);
\coordinate[label={[below]\textbf{$\mathbf{53}$}}] (a) at (-0.5,-4.3);
\coordinate[label={[below]\textbf{$\mathbf{49}$}}] (a) at (0.5,-4.3);
\coordinate[label={[below]\textbf{$\mathbf{40}$}}] (a) at (1.5,-4.3);
\coordinate[label={[below]\textbf{$\mathbf{48}$}}] (a) at (2.5,-4.3);
\coordinate[label={[below]\textbf{$\mathbf{54}$}}] (a) at (-2.5,-5.3);
\coordinate[label={[below]\textbf{$\mathbf{54}$}}] (a) at (-1.5,-5.3);
\coordinate[label={[below]\textbf{$\mathbf{54}$}}] (a) at (-0.5,-5.3);
\coordinate[label={[below]\textbf{$\mathbf{52}$}}] (a) at (0.5,-5.3);
\coordinate[label={[below]\textbf{$\mathbf{55}$}}] (a) at (1.5,-5.3);
\coordinate[label={[below]\textbf{$\mathbf{58}$}}] (a) at (2.5,-5.3);
\end{tikzpicture}}}
     \caption{(a) A PET detector ring (radius 450 mm) of the size of a  typical human body PET scanner and the geometry ($388 \times 388$ $mm^{2}$) of an scanned object;\ (b) pixelized approximation of the chosen activity distribution with the number of coincident events generated (mentioned inside) from each of them.}
  \label{fig4}
\end{figure*}

A PET detector ring of radius 450 mm, similar to the size of a typical human PET scanner, was chosen. Figure~\ref{fig4}(a) presents a nonuniform activity distribution ($388 \times 388\ mm^{2}$). Figure~\ref{fig4}(b) shows the activity distribution consisting of $1 \times 1 \ mm^{2}$ pixels with annihilation point at its below-left corner. The number of coincident events generated from a pixel is the number contained in that pixel. We achieved simplicity in simulation by assuming a 2D geometry. Indeed, our assumption did not undermine the applicability of the proposed technique to 3D-PET, as explained in Sec.~\ref{sec:level2_1_2}, and further discussed in Sec.~\ref{sec:level3_2}.
\par

We assumed that a uniform attenuating medium was filled inside the entire space of the PET detector ring. Although, it was an unrealistic assumption as attenuating medium should only be inside the scanned object boundary, we understood that it would not impose any fundamental limitation on the applicability of our proposed technique in a realistic situation. In fact, the assumption was made to simplify the geometry of simulation. Without such an assumption, we needed to perform the Virtual extrapolation from the boundary of the scanned object, located at points $A^{'}$, $B^{'}$, instead from the points ($A$ and $B$ in Fig.~\ref{fig3}(a) at the PET ring. Also, we needed to fill the Virtual attenuating medium outside the scanned object, rather than filling outside the PET ring.
\par

We tuned the attenuation coefficient of the medium so that the relative fractions of USE, SSE, and MSE resembled a typical case of PET scan. We found that for attenuation coefficient $\mu=\ 0.01\ cm^{-1}$ (compared to $\mu$ for water = $0.096 \ cm^{-1}$), the percentages of USE, SSE, and MSE were $41.84\%$, $46.45\%$, and $11.71\%$, respectively, when the total number of simulated coincident events was $10\,824\,391\ (\sim 10^{7})$. The total scatter fraction (SSE+MSE) was high (58\%) for a typical 2D-PET scan. Nevertheless, we accepted it because the feasibility of our model was not limited to a 2D-PET system; we could as well apply it to a 3D-PET system with genuinely high fraction of scattering. The reason of getting high scatter fraction, even using much lower value of $\mu$ than for water, was because the attenuating medium filled the entire space inside the PET ring, not just inside the phantom. The high SSE fraction turned out to be favorable for our proposed data analysis algorithm, described in Secs.~\ref{sec:level4}, \ref{sec:level5} and \ref{sec:level6}, which is in its early developmental stage.\par

To generate the coincident events one by one from the pixels shown in Fig.~\ref{fig4}(b), a uniform probability density function (PDF) between $0\degree$ to $180\degree$ was applied, giving randomly the direction of emission of two coincident photons. In the next step, the exponential probability density function (Eq.~\ref{equ1}) was used to randomly decide lengths from the annihilation point $O$ (Fig.~\ref{fig3}(a)) of two independent single scattered points. By recognizing the number of real scattering points -- zero for USE, one for SSE, and more than one for MSE -- inside the detector ring for a coincident event, we classified all the events. Ignoring MSE, we finally used the Klein-Nishina PDF for unpolarized photons \cite{bailey2005positron} for each SSE, defined for the range $-\pi\leq\theta<\pi$, to randomly decide the scattering angle (Eq.~\ref{equ3}).\par

\begin{eqnarray}
p(\theta)=0.2237 \left \{ \frac{3-3\cos\theta+3\cos^{2}\theta-\cos^{3}\theta}{(2-\cos\theta)^{3}} \right \}
\label{equ3}
\end{eqnarray}

\subsection{\bf{Transforming PET scan data through Virtual extrapolation}}
\label{sec:level3_2}

\begin{figure}[htb] 
\centering 
\includegraphics[width=0.5\textwidth,height=0.45\textheight,keepaspectratio]{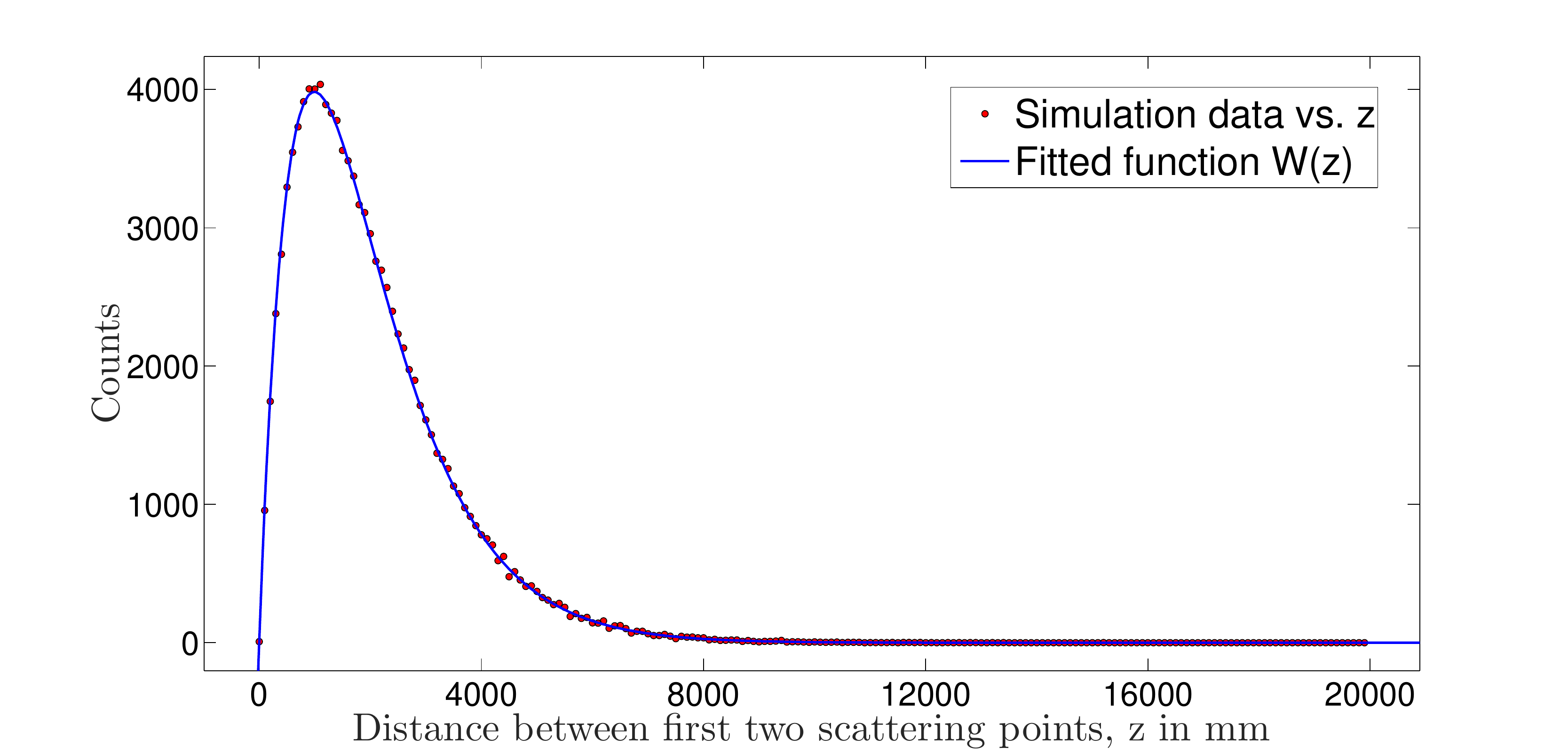}
\caption{Transformed PET scan data fitted with the PDF of Eq.~\ref{equ2} for validating Virtual extrapolation technique in the case of precise timing information.}
\label{fig5}
\end{figure}

We carried out the transformation of coincident events of the simulated PET data (described in the previous section) using the Virtual extrapolation technique. The transformed data were fitted with the PDF ($W(z)$) (Eq.~\ref{equ2}) with some constant multiplying factor, presented in Fig.~\ref{fig5}. Once transformed, all events became the data set in one variable $z$. The variable $z$ is the distance between two scattering points - either real or virtual - of two coincident photons in the infinitely extended uniform attenuating medium. It is noteworthy that sometimes the value of $z$ could be higher than the diameter of the PET ring, indeed a realistic situation with non-zero probability. When visualizing the original data for these high-$z$ values, the corresponding event consisted of either one or no Compton scattering out of the two coincident photons. The excellent overlap of the data with the mathematical function (Eq.~\ref{equ2}) proves the validity of Virtual extrapolation technique (Fig.~\ref{fig5}), ensuring the correctness of the transformed data.\par

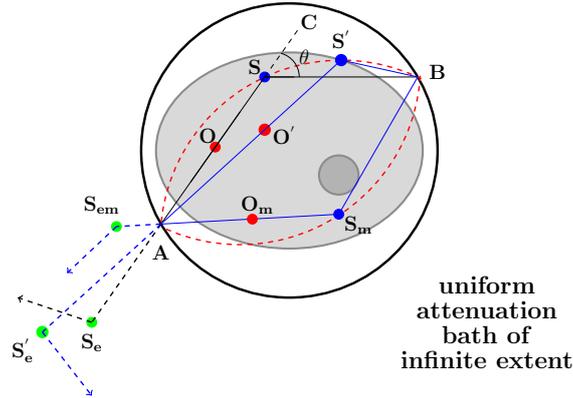
\begin{figure}[htb]
  \centering
  \scalebox{0.65}{\begin{tikzpicture}
\draw[line width=0.5mm,solid] (0,0) circle (3cm);
\filldraw[color=gray!90,fill=gray!30,very thick](0,0) ellipse (2.7 cm and 2.0 cm);
\filldraw[color=gray!90,fill=black!30,very thick](1.0,-0.5) circle (0.4cm);
\draw [thick,red,dashed,domain=63.719:174.988] plot ({1.049+3.661*cos(\x)}, {-1.817+3.661*sin(\x)});
\draw [thick,red,dashed,domain=-114.0:-3.0] plot ({-1.005+3.661*cos(\x)}, {1.740+3.661*sin(\x)});
\draw[thin,red,fill=red,solid] (-1.5,0.07) circle (0.1cm);
\draw[thin,blue,fill=blue,solid] (-0.5,1.5) circle (0.1cm);
\draw(-2.598,-1.500) -- (-0.5,1.5);
\draw(-1.5,0.07) -- (-2,-0.645);
\draw[thin,black,solid](-0.5,1.5) -- (2.598,1.500);
\draw(-1.5,0.07) -- (-1.0,0.785);
\draw(-0.5,1.5) -- (0.1,1.5);
\draw[thin,green,fill=green,solid] (-4,-3.505) circle (0.1cm);
\draw[thick,black,dashed](-2.598,-1.5) -- (-4,-3.505);
\draw[thick,black,dashed,->](-4,-3.505) -- (-5.5,-3);
\coordinate[label={[below]$\mathbf{S_{e}}$}] (c) at (-4,-3.6);
\draw[thin,blue,fill=blue,solid] (1,-1.3) circle (0.1cm);
\draw[thin,blue,solid](1,-1.3) -- (-2.598,-1.5);
\draw[thin,blue,solid](1,-1.3) -- (2.598,1.5);
\draw[thin,red,fill=red,solid] (-0.75,-1.4) circle (0.1cm);
\draw[thin,green,fill=green,solid] (-3.5,-1.551) circle (0.1cm);
\draw[thick,blue,dashed](-2.598,-1.5) -- (-3.5,-1.551);
\draw[thick,blue,dashed,->](-3.5,-1.551) -- (-4.5,-2.5);
\coordinate[label={[below]$\mathbf{S^{'}_{e}}$}] (c) at (-5.4,-3.7);
\draw[line width=0.5mm,,red,fill=red,solid] (-0.5,0.424) circle (0.1cm);
\draw[thin,,blue,solid](-2.598,-1.500) -- (1.049,1.844);
\draw[thin,,blue,solid](1.049,1.844) -- (2.598,1.500);
\draw[line width=0.5mm,,blue,fill=blue,solid] (1.049,1.844) circle (0.1cm);
\draw[line width=0.5mm,,green,fill=green,solid] (-5,-3.703) circle (0.1cm);
\draw[thick,,blue,dashed](-2.598,-1.5) -- (-5,-3.703);
\draw[thick,,blue,dashed,->](-5,-3.703) -- (-4,-5);
\coordinate[label={[below]\textbf{A}}] (a) at (-2.598,-1.8);
\coordinate[label={[below]\textbf{B}}] (a) at (3,1.9);
\coordinate[label={[below]\textbf{$\mathbf{S_{m}}$}}] (a) at (1.4,-1.2);
\coordinate[label={[below]\textbf{$\mathbf{O_{m}}$}}] (a) at (-0.65,-0.8);
\coordinate[label={[below]\textbf{O}}] (a) at (-1.65,0.6);
\coordinate[label={[below]\textbf{S}}] (a) at (-0.7,2.0);
\coordinate[label={[below]$\mathbf{O^{'}}$}] (c) at (-0.1,0.65);
\coordinate[label={[below]$\mathbf{S^{'}}$}] (c) at (1.049,2.6);
\coordinate[label={[below]$\mathbf{S_{em}}$}] (c) at (-3.8,-0.8);
\draw[thin,black,dashed](-0.5,1.5) -- (0.2,2.501);
\draw (0.2,1.5) arc (0:75:5mm);
\coordinate[label={[below]\textbf{$\theta$}}] (e) at (0.3,2.2);
\coordinate[label={[below]\textbf{C}}] (e) at (0.4,2.9);
\node[align=center] at (4,-3.5) {\large{\textbf{uniform}}\\ \large{\textbf{attenuation}}\\ \large{\textbf{bath of}}\\ \large{\textbf{infinite extent}}};
\end{tikzpicture}}
  \caption{Locus (dashed in red) of the scattering point $S$ with annihilation point $O$ and its virtually extrapolated point at $S_e$; the detection points are located at $\mathbf{B}$ (scattered photon) and $\mathbf{A}$ (unscattered coincident photon).  Similarly, another possible scattering point $S^\prime$ on the same locus, with annihilation point $O^\prime$ and virtually extrapolated point $S^\prime_e$, is also shown. For completeness, another mirror symmetric locus (dashed in red) -- with scattering point $S_m$, annihilation point $O_m$ and virtually extrapolated point $S_{em}$ -- is traced.}
  \label{fig6}
\end{figure}

Figure~\ref{fig6} shows a SSE with scattering at $S$ and annihilation point at $O$, and two photon detection points at $A$ and $B$. Knowing these detection points from the original data and the scattering angle ($\theta$) from Eq.~\ref{equ4}, we could identify the locus of scattering points as shown in dotted red curve $\widetilde{ASB}$. The loci are the arcs of two equal radius circles in 2D-PET \cite{conti2012reconstruction}. Similarly, we can get the surface of a prolate spheroid \cite{hemmati2017compton} in 3D-PET. An extra coordinate  -- azimuthal angle ($\phi$) describing the rotation around polar axis passing through the points $A$ and $B$ -- would be needed to specify the scattering point $S$ in 3D-PET. The scattering angle ($\theta$) of a photon of energy E was calculated from the Compton kinetic equation,
\begin{eqnarray}
\theta={\cos}^{-1}(2-\frac{511}{E}).
\label{equ4}
\end{eqnarray}

As seen from Fig.~\ref{fig6}, if the original scattering point `$S$' was known, the line of response (LOR) could be retraced through `$S$' and $A$ (the unscattered photon detection point). The retraced LOR information could then be used to add that event in the image reconstruction.\par

\begin{table*}[htb]
\centering
\footnotesize
\begin{tabular}{@{}||c|c|c|c|c|c|c|c|c}
\hline
use-1&sse-1&sse-2&mse-1&use-2&use-3&sse-3&use-4&---\\ [0.5ex] 
 \hline\hline

  	    900.5  & 543.2 & 639.0 & 124.6 & 723.3 & 983.3 & 739.3 & 1278.2 \\
                             & 738.0 & 904.9 &               &                &              & $\textcolor{blue}{839.9^{*}}$  &\\
                             & $\textcolor{blue}{837.7^{*}}$ & 725.3 &               &                &              & 633.9  &\\
                  & 764.9 & 836.8 &               &                &              & 1073.3&\\
                             & 563.0 & $\textcolor{blue}{1389.3^{*}}$&             &                 &              & 907.3  &\\
                             &  ---         &  ---          &              &                 &              &      ---       &\\
                             & ---          &   ---         &              &                 &              &       ---      &\\
 \hline
\end{tabular}\\
\caption{\label{eventbyeventtable}Transformed PET scan data in variable $z$ (mm) written event by event with multiple possibilities for the SSE events.  In all the SSE columns, one value corresponds to the real scattering point (marked as * in blue).}\end{table*}
\normalsize

Since the real scattering point was unknown, we needed to do the Virtual extrapolation for each possible scattering point on the circular locus $\widetilde{ASB}$. For instance, $S^\prime$ is yet another possible scattering point with corresponding Virtually extrapolated point $S^\prime_e$ and anihillation point $O^\prime$. To reduce the computational load, we considered only discrete scattering points with separation of 1 mm arc length on the locus. The separation could be increased (or decreased) with less (or more) computational load at the cost of reduced (or increased) accuracy in recognizing the original scattering point.\par

Extending the idea to a 3D-PET system, we would need to perform the Virtual extrapolation for all possible scattering points on the surface of a prolate spheroid.

\subsection{\bf{Collective difference property}}
\label{sec:level3_3}

Once we obtained the transformed PET data by Virtual extrapolation, we used that data set for all our further analysis. It is needless to mention again that our transformed data became one-parameter list mode data in variable $z$ (mm). A part of the data is presented in the tabular format (Table~\ref{eventbyeventtable}). In the table, USE-1, USE-2,.... denote first, second,...., respectively, the unscattered events as happened sequentially in the simulation environment; similarly, SSE-1, SSE-2,.... are the single scattered events, and MSE-1, MSE-2.... are the multi-scattered events. Columns corresponding to SSE events contain multiple values, one corresponding to the original (real) scattering point, and remaining to other possible scattering points artificially placed on the locus (circular arc in Fig.~\ref{fig6}), as described in the previous section. Our task of data analysis was to find the original scattering points, marked as ``*'' (in blue) in the table.\par

We proposed a data analysis algorithm which enabled us to find the real scattering point. We succeeded because of a unique `collective difference property'. By collective difference, we mean a recognizable difference in the transformed data, only achieved by comparing the data set containing all the real (original) scattering points, with those including false scattering points artificially placed on the locus. In other words, the collective difference happened due to special overall behavior of the transformed data set representing the original scattering points (starred ``*'' (in blue) in Table~1) -- following the distribution function $W(z)$ (Eq.~\ref{equ2}) -- in comparison with the remaining data sets.
\par

The collective property originates from both the components of travel path of a coincident photon pair. For the travel path lying inside the detector ring corresponding to a real (original) scattering point, we obtained the distance by a true random value generated through Monte Carlo (MC) simulation. In other words, the distance $\mathbf{OS}$ in Fig.~\ref{fig6} -- for the annihilation point $\mathbf{O}$ and the actual scattering point $\mathbf{S}$ -- obeyed the single scattered attenuating path length distribution (Eq.~\ref{equ1}). Hence, it was a correct random value. However, for false scattering points $\mathbf{S^{'}}$, the length $\mathbf{O^{'}S^{'}}$ could be arbitrary. To note again, we placed the false scattering points ($\mathbf{S^{'}}$) on the arc of circle (mentioned in the previous section) such that the scattering angle ($\theta$) was the same as for $\mathbf{S}$. Also, the path length $\mathbf{OSB}$ was same as $\mathbf{O^{'}S^{'}B^{'}}$. The distances like $\mathbf{{O}^{'}{S}^{'}}$ did not obey the probability distribution of single scattered attenuating path lengths (Eq.~\ref{equ1}), therefore could be false random values. Further, for the partner coincident photon detected at $\mathbf{A}$ -- corresponding to only the real scattering point $\mathbf{S}$ -- the distance of `virtual' travel path outside the PET ring (point $\mathbf{S_e}$) was allocated correctly through Virtual extrapolation. As a result, both the components of photon travel paths got the correct random values. It is worth mentioning that a part of the collective diference property, \textit{i.e.}, the path lengths expected to follow W(z) (Eq.~\ref{equ1}), was already checked through correct fitting of the transformed data in Sec.~\ref{sec:level3_2}. Further, both TOF-PET and non-TOF-PET comply to the collective difference property.

\subsubsection{An implicit assumption}
\label{sec:level3_4}

We earlier claimed that the only data value corresponding to the real scattering point ($\mathbf{S}$) obeys the probability density function (PDF) of Eq.~\ref{equ2} due to collective difference property. But in reality, there exists another data value corresponding to the false scattering point ($\mathbf{S_{m}}$) -- positioned at the mirror symmetric location of $\mathbf{S}$ (Fig.~\ref{fig6}) -- which mimics the same collective property because of the same distances $\overline{\mathbf{AS}}$ and $\overline{\mathbf{AS_{m}}}$. In case of a 3D-PET system, there are more than one mirror symmetric points, and they form a circle on the surface of a prolate spheroid in the equatorial plane. So, whatever may be the data analysis algorithm, we cannot distinguish the two events corresponding to $\mathbf{S}$ and $\mathbf{S_{m}}$ by using only the idea of difference in collective behavior.\par

In the present work, we have ignored the problem of mirror symmetry by choosing only $\mathbf{S}$ (not $\mathbf{S_{m}}$), because $\mathbf{S}$ could be a known scattering point for the simulated PET data. Intuitively, the two mirror symmetric points could be distinguished easily for a non-uniform phantom, which is planned for our future work.\par 

\subsubsection{Finite time and energy resolution}
\label{sec:level3_3_1}

\begin{figure}[htb]
  \centering
  \scalebox{0.65}{\begin{tikzpicture}
\draw[line width=0.5mm] (0,0) circle (3cm);
\filldraw[color=gray!90,fill=gray!30,very thick](0,0) ellipse (2.7 cm and 2.0 cm);
\filldraw[color=gray!90,fill=black!30,very thick](1.0,-0.5) circle (0.4cm);
\draw [thick,red,dashed,domain=63.719:174.988] plot ({1.049+3.661*cos(\x)}, {-1.817+3.661*sin(\x)});
\draw [thick,blue,dashed,domain=80.0:160.0] plot ({1.7+4.5*cos(\x)}, {-2.9+4.5*sin(\x)});
\draw [thick,blue,dashed,domain=48.719:187.988] plot ({0.5+3.15*cos(\x)}, {-0.8+3.15*sin(\x)});
\draw [thick,blue,dashed,domain=-30.0:85.0] plot ({-2.9+3.0*cos(\x)}, {-1.5+3.0*sin(\x)});
\draw[red,fill=red!100] (-1.3,1.0) circle (0.08cm);
\draw[blue,fill=blue!100] (-0.9,0.75) circle (0.08cm);
\draw[blue,fill=blue!100] (-1.85,1.3) circle (0.08cm);
\coordinate[label={[below]$\mathbf{S_{1}}$}] (a) at (-1.85,2.0);
\coordinate[label={[below]$\mathbf{S_{2}}$}] (a) at (-0.4,1.00);
\coordinate[label={[below]$\mathbf{S}$}] (a) at (-1.3,1.65);
\coordinate[label={[below]$\mathbf{O}$}] (a) at (-1.4,0.3);
\draw[red,fill=red!100] (-1.7,0.2258) circle (0.08cm);
\draw(-2.598,-1.500) -- (-1.3,1.0);
\draw(-1.3,1.0) -- (2.68,1.48);
\coordinate[label={[below]\textbf{A}}] (a) at (-2.9,-1.500);
\coordinate[label={[below]\textbf{B}}] (b) at (3.0,1.9);
\end{tikzpicture}}
  \caption{A single scattered coincident event (SSE) detected at points $\mathbf{A}$ and $\mathbf{B}$, where $\mathbf{S}$ is the original scattering point and $\mathbf{O}$ the annihilation point; $\mathbf{S}$ lies on the true locus circle $\widetilde{\mathbf{ASB}}$, but there exist possible misplaced locus circles at $\widetilde{\mathbf{AS_{1}B}}$ and $\widetilde{\mathbf{AS_{2}B}}$. Points $\mathbf{S_{1}}$ and $\mathbf{S_{2}}$ can be recognized as true scattering points on the misplaced locus circles such that $\overline{\mathbf{AS_{1}}}=\overline{\mathbf{AS_{2}}}=\overline{\mathbf{AS}}$.}  \label{fig7}
\end{figure}
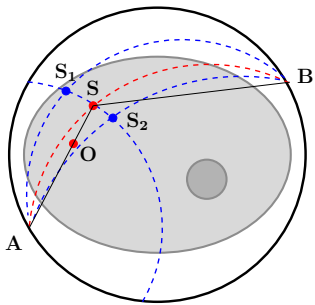

~~~So far, we have assumed perfect timing and energy information about the coincident photons in describing our model. Once finite detector resolutions are invoked, we expected some inaccuracies in our model prediction, questioning the validity of our approach. In the case of detectors with finite timing resolution, annihilation points can drift from their actual position. It is needless to mention that the aim of our present work is to find the original scattering points, not the annihilation points. The fact that the model is expected to work for a non-TOF-PET system because of the concept of length compensation, mentioned in Sec.~\ref{sec:level2_2} and elaborated in the \ref{appen1}, the annihilation point $\mathbf{O}$ (Fig.~\ref{fig7}) could be arbitrarily chosen on the line $\mathbf{AS}$. The transformed data corresponding to the real scattering points should still obey $W(z)$ distribution (Eq.~\ref{equ2}). Consequently, the collective difference property should still remain preserved. However, a degraded time resolution could indirectly affect the performance by increasing the amount of random coincidences.
\par

Due to the finite energy resolution of detectors and error propagation, the scattering angle ($\theta$) (Eq.~\ref{equ4}) can be uncertain. As a result, the position of the true locus circle (red dashed curve in Fig.~\ref{fig7}) gets misplaced somewhere in the vicinity (dashed curves in blue). The amount of shift -- causing blurring in the locus circle -- depends on the energy resolution of detectors. Since the position of the annihilation point $\mathbf{O}$ is irrelevant, the distance $\overline{\mathbf{AS}}$ is the measure of trueness. However, there exist many such distances, \textit{e.g.}, $\overline{\mathbf{AS_{1}}}$, $\overline{\mathbf{AS_{2}}}$,.... (Fig.~\ref{fig7}) to assign `true' scattering points $S_1$, $S_2$,...., respectively. Interestingly, these misplaced points and real scattering point ($\mathbf{S}$) lie on the locus of yet another circle, termed as `Symmetric Second Circle'. We named it Second circle considering the earlier one as First circle containing all possible scattering points $\mathbf{S}$ and $\mathbf{S^{'}}$ (Fig.~\ref{fig6}). Here, `symmetric' means the points on it are indistinguishable as far as the proposed model is concerned. In other words, the proposed model cannot recognize the true scattering point from all other points on the second symmetric circle.  In essence, the success of the present model depends on using detectors with very good energy resolution, \textit{e.g.}, LaBr$_{3}$:Ce. In this context, it is worth noting that even if the proposed model is unable to select out true scattering point from all possible ones on the Symmetric Second circle, the recognized scattering point remains close to the real one ($\mathbf{S}$ in Fig.~\ref{fig7}) for detectors with good energy resolution. Further, the indistinguishable points are expected to be recognized for a non-uniform medium -- a study we plan to carry out in future as an extension of present work.

\subsection{\bf{Data analysis}}
\label{sec:level4}

We proposed a data analysis algorithm which is presented in a general term at first. Later in Secs.~\ref{sec:level5} and \ref{sec:level6}, we give a step by step account of the proposed algorithm for the three cases -- ideal, finite time and finite energy resolutions.\par

From Table~1, we created all possible data combinations -- each consisting of one value taken from every column -- which turned out to be large in number because of many different $z$-values in the SSE columns. Among all the combinations, only one was unique containing all the real scattering data values from the SSE columns of Table~1 (starred ``*'' in blue), which followed the $W(z)$ distribution (Eq.~\ref{equ2}). Our task was to find that unique combination among all possibilities using the `collective difference property'. However, the total number of possible combinations became too high ($200^{10^{5}}$), if we assumed $200$ possible data values for each SSE column in Table~1, and the total number of SSE columns to be $10^{5}$. In fact, the number of data values in a particular SSE column depends on the arc length of the circular locus and the choice of the separation length (1 mm in our case) between the discrete scattering points (Sec.~\ref{sec:level3_2}). To avoid handling such a huge data set, we used the concept of random sampling \cite{casella2002statistical} -- a true representation of the whole data set. So, instead of dealing with the complete transformed PET data set (Table~1), a random sample of small size, possessing the same collective difference property as the whole data, was chosen for the data analysis. We, thus, generated a data set containing a small number of combinations to be checked. Since the list mode data collection is possible in a real PET scan \cite{nichols1999continuous}, random sampling was possible by `sequential sampling'.\par

To quantify how close the chosen random data set was in obeying $W(z)$ (Eq.~\ref{equ2}), we defined a sampling statistic \cite{ross2017introductory} `$\Delta$', given as, 

\begin{eqnarray}
\Delta=\sum_{i=1}^{N}\left ( \frac{f_{i}}{n}-w_{i} \right )^{2}.
\label{equ5}
\end{eqnarray}

Here, $f_i$ is the relative frequency of the transformed PET data ($z$-values) for the $i^{th}$ bin, and $w_i$ is the corresponding theoretical value from Eq.~\ref{equ2}. The sample size is $n$, and $N$ is the total number of bins chosen for the whole spectrum of double scattering distances ($z$) ranging from $0$ to $2200$ mm. It is noteworthy that the entire formulation of the mean and variance for the sample statistic `$\Delta$', presented in \ref{appen2}, played a vital role in the final stage of our data analysis. The derived results (Eqs.~\ref{equA1} and \ref{equA2}) are as follows:\par

\begin{equation}
\begin{aligned}
Mean: f(n)=\frac{1}{n}\sum_{i=1}^{N}\left ( w_{i}-w_{i}^{2}\right ),
\label{equ6}
\end{aligned}
\end{equation}

\begin{equation}
\begin{aligned}
&Variance: V(n)=\left ( \frac{2}{n^{2}}-\frac{6}{n^3} \right )\sum_{i=1,j=1}^{N}w_{i}^{2}w_{j}^{2}\\
&+\left ( \frac{4}{n^{3}}\right )\sum_{i=1,j=1}^{N}w_{i}^{2}w_{j}- \left (\frac{1}{n^{3}} \right )\sum_{i=1,j=1}^{N}w_{i}w_{j}.
\end{aligned}
\label{equ7}
\end{equation}

\begin{figure}[htb]  
\centering
\includegraphics[width=0.45\textwidth, height=0.45\textheight,keepaspectratio]{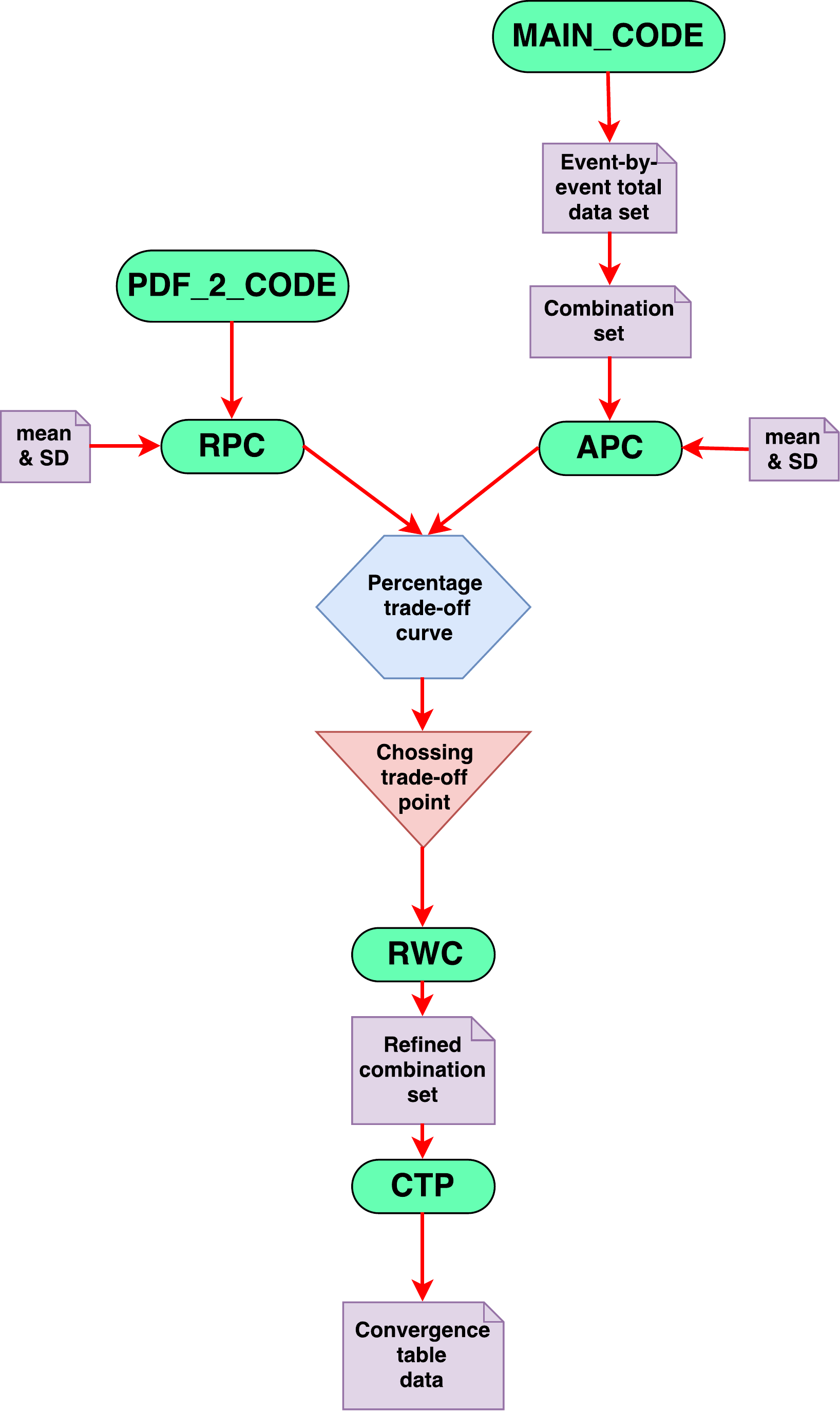}
\caption{Flowchart for the data analysis steps; the acronyms are described in the text.}
\label{fig8}
\end{figure}

Figure~\ref{fig8} presents the flow chart for our data analysis code. There are two branches, one for generating the simulated transformed PET data and another for finding the theoretical probability density function. Later these two branches are merged to find the correct data value corresponding to the original scattering point for each SSE. The details are described in the following two paragraphs. \par

\begin{figure}[htb]  
\centering
\includegraphics[width=0.53\textwidth, height=0.53\textheight,keepaspectratio]{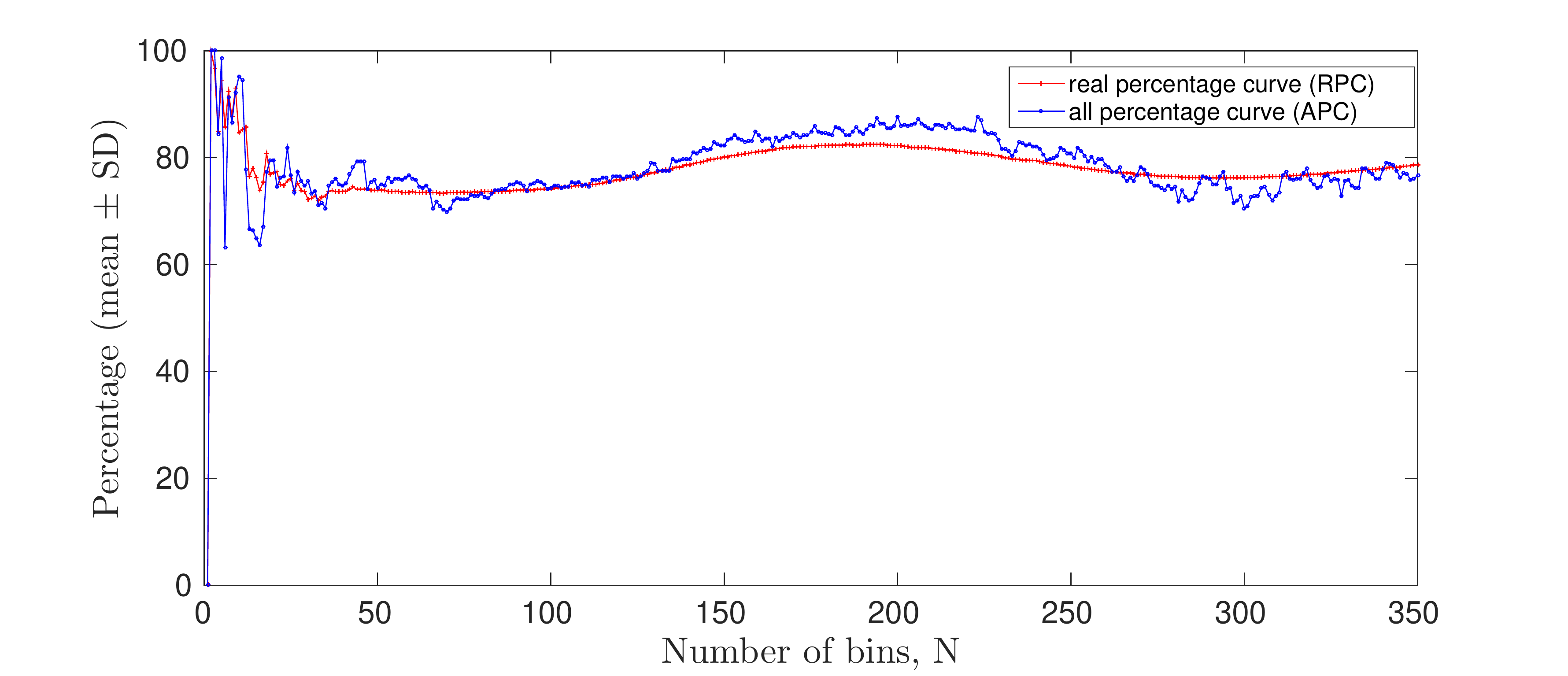}

\caption{Percentage trade-off curves.}
\label{fig9}
\end{figure}

Through the code PDF\_2\_CODE (Fig.~\ref{fig8}), we first generated $\sim 10^{7}$ event-by-event sequential data ($z$-values) corresponding to the PDF of Eq.~\ref{equ2} using Monte Carlo simulation. The random samples of size `\textit{n}' were sequentially drawn from the data. We calculated `$\Delta$' (Eq.~\ref{equ5}), its mean (Eq.~\ref{equ6}) and standard deviation (SD) (Eq.~\ref{equ7}) for the equal-sized bins. Many \textit{N} values (=1,2,3,...,350) were tried out. Our objective was to determine how many percentage of random samples $\Delta$ lied between mean $\pm$ SD. Figure~\ref{fig9} shows a typical plot of these percentages as a function of \textit{N} for the sample size \textit{n}=6, named as `\textbf{R}eal \textbf{P}ercentage \textbf{C}urve' generated using a C++ code (RPC). Through the other branch in Fig.~\ref{fig8}, we created `\textbf{A}ll \textbf{P}ercentage \textbf{C}urve' (APC) similar to RPC. The only difference was in using the original data, \textit{i.e.}, for APC we used the simulated event-by-event transformed PET data (Table~\ref{eventbyeventtable}) to build the combination data sets for the first \textit{n} events. Rest of the procedure was the same as that for RPC. For instance, first few combination sets for \textit{n}=6 were as follows:

\begin{center}
[ (900.5, 543.2, 639.0, 124.6, 723.3);\\
  (900.5, 543.2, 904.9, 124.6, 723.3);\\
  ..............;\\
  (900.5, 738.0, 639.0, 124.6, 723.3);\\
  (900.5, 738.0, 904.9, 124.6, 723.3);\\
  ................].
\end{center}

With both RPC and APC in hand, we aimed to find a refined combination data set to identify the actual scattering point. To begin with, we needed to find an appropriate value of \textit{N}. A high value of RPC meant that the combinations for which all the data values, rightly obtained using W($z$) (Eq.~\ref{equ2}), had a high chance of remaining inside their two bounds of mean $\pm$ SD. To find the value of \textit{N}, we utilized APC when most of the combinations from the set got rejected. So, the best choice of \textit{N} from Fig.~\ref{fig9} was a trade-of-point when the value for RPC was as high as possible with minimum possible value for APC. In our further data analysis, we consistently found the desired results for the medium values of $N$ ($N>$50), avoiding the initial high fluctuation region. Our choice of \textit{N}, however, depended on the sample size \textit{n}. We used the combination-set file as input to RWC (`\textbf{R}eject \textbf{W}ild \textbf{C}ombinations') program (Fig.~\ref{fig8}), rejecting all the combinations for which `$\Delta$' was outside the acceptable range of mean $\pm$ SD. We thus obtained a file of refined combination set. Finally, the frequencies were assigned for the repetitive occurrence of the data values for each SSE using the program CTP (`\textbf{C}onvergence \textbf{T}able \textbf{P}roduction'). The aim was to identify the real scattering point by locating the data value corresponding to the highest frequency.\par

We used a high-performance computer \cite{yuvaindia} for running all the data generating Monte Carlo C++ codes (MAIN\_CODE and PDF\_2\_CODE) and the data analysis codes (RPC, APC, RWC, and CTP).  

\section{\large Results}
\label{sec:level5}

The following sections present the results of our data analysis one by one for the three cases -- ideal, finite time, and energy resolutions. We first handled the ideal cases taking small samples (Sec.~\ref{sec:level4}) -- chosen with certain restrictions, discussed in Sec.~\ref{sec:level6}, on the parameter values - to obtain meaningfully good results.

\subsection{\bf{Ideal energy and time resolution}}
\label{sec:level5_1}

\begin{table*}[htb!]
\centering
\begin{minipage}[b]{0.40\linewidth}
\centering
\begin{tabular}{||c | c | c||}
\hline
\scriptsize
No. & z (mm) & Frequency  \\ [0.5ex] 
\hline\hline
1	&	2166.70	&	873787	\\	\hline
2	&	1015.25	&	727567	\\	\hline
3	&	1788.94	&	863171	\\	\hline
4	&	1171.01	&	856961	\\	\hline
5	&	1342.66	&	856961	\\	\hline
6	&	1204.50	&	856961	\\	\hline
7	&	3075.88	&	911716	\\	\hline
8	&	1056.82	&	856961	\\	\hline
9	&	1842.72	&	863171	\\	\hline
10	&	1088.68	&	856961	\\	\hline
11	&	819.03	&	727567	\\	\hline
12	&	729.01	&	727567	\\	\hline
13&$\textcolor{red}{3480.63^{*}}$&$\textcolor{red}{937724^{*}}$	\\	\hline
\end{tabular}
\end{minipage}
\begin{minipage}[b]{0.45\linewidth}
\centering
\begin{tabular}{||c | c | c||} 
\hline
No. & z (mm) & Frequency  \\ [0.5ex] 
\hline\hline
1	&	1156.63	&	899157	\\	\hline
2	&	2474.67	&	$\textcolor{red}{990896^{*}}$	\\	\hline
3	&	$\textcolor{red}{2164.42^{*}}$	&	977170	\\	\hline
4	&	817.30	&	882108	\\	\hline
5	&	749.96	&	882108	\\	\hline
6	&	568.55	&	909828	\\	\hline
7	&	475.96	&	909828	\\	\hline
8	&	421.39	&	909828	\\	\hline
9	&	871.41	&	882108	\\	\hline
10	&	639.40	&	909828	\\	\hline
11	&	839.76	&	882108	\\	\hline
12	&	943.34	&	882108	\\	\hline
\end{tabular}
\end{minipage}

\caption{Analyzed results for the data samples with cardinality 13 (left) and cardinality 12 (right), for the case of ideal energy and timing informations; see details in the text.}
\label{table2}
\end{table*}

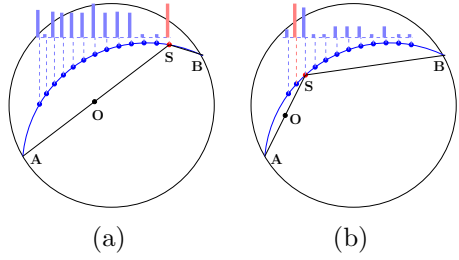
\begin{figure}[htp]
  \centering
  \subfigure[]{\scalebox{0.45}{\begin{tikzpicture}
\draw[line width=0.2mm] (0,0) circle (3cm); 
\draw [thick,blue,domain=63.719:174.988] plot ({1.049+3.661*cos(\x)}, {-1.817+3.661*sin(\x)});
\draw[dashed,blue!50] (-2.072,2.0) -- (1.735,2.0); 
\draw[red,fill=red!100] (1.685,1.788) circle (0.06cm); 
\path [fill=red!50,draw=red!50,line width=1mm]
(1.635,2.0) -- (1.635,3.0); 
(1.735,2.0) -- (1.735,3.0);
\draw[dashed,red!50] (1.685,1.788) -- (1.685,2.0);
\draw[blue,fill=blue!100] (1.317,1.834) circle (0.06cm); 
\path [fill=blue!50,draw=blue!50,line width=1mm]
(1.317,2.0) -- (1.317,2.1); 
(1.267,2.0) -- (1.267,2.1);
\draw[dashed,blue!50] (1.317,1.834) -- (1.317,2.0);
\draw[blue,fill=blue!100] (0.947,1.843) circle (0.06cm); 
\path [fill=blue!50,draw=blue!50,line width=1mm]
(0.897,2.0) -- (0.897,2.1); 
(0.997,2.0) -- (0.997,2.1);
\draw[dashed,blue!50] (0.947,1.843) -- (0.947,2.0);
\draw[blue,fill=blue!100] (0.577,1.814) circle (0.06cm); 
\path [fill=blue!50,draw=blue!50,line width=1mm]
(0.527,2.0) -- (0.527,2.732); 
(0.627,2.0) -- (0.627,2.732);
\draw[dashed,blue!50] (0.577,1.814) -- (0.577,2.0);
\draw[blue,fill=blue!100] (0.213,1.747) circle (0.06cm); 
\path [fill=blue!50,draw=blue!50,line width=1mm]
(0.163,2.0) -- (0.163,2.762); 
(0.263,2.0) -- (0.263,2.762);
\draw[dashed,blue!50] (0.213,1.747) -- (0.213,2.0);
\draw[blue,fill=blue!100] (-0.143,1.645) circle (0.06cm); 
\path [fill=blue!50,draw=blue!50,line width=1mm]
(-0.193,2.0) -- (-0.193,2.732);   
(-0.093,2.0) -- (-0.093,2.732);
\draw[dashed,blue!50] (-0.143,1.643) -- (-0.143,2.0);
\draw[blue,fill=blue!100] (-0.487,1.506) circle (0.06cm); 
\path [fill=blue!50,draw=blue!50,line width=1mm]
(-0.537,2.0) -- (-0.537,2.999); 
(-0.437,2.0) -- (-0.437,2.999);
\draw[dashed,blue!50] (-0.487,1.506) -- (-0.487,2.0);
\draw[blue,fill=blue!100] (-0.815,1.334) circle (0.06cm);  
\path [fill=blue!50,draw=blue!50,line width=1mm]
(-0.865,2.0) -- (-0.865,2.732);
(-0.765,2.0) -- (-0.765,2.732);
\draw[dashed,blue!50] (-0.815,1.334) -- (-0.815,2.0);
\draw[blue,fill=blue!100] (-1.124,1.130) circle (0.06cm);  
\path [fill=blue!50,draw=blue!50,line width=1mm]
(-1.174,2.0) -- (-1.174,2.732);
(-1.074,2.0) -- (-1.074,2.732);
\draw[dashed,blue!50] (-1.124,1.130) -- (-1.124,2.0); 
\draw[blue,fill=blue!100] (-1.410,0.895) circle (0.06cm);  
\path [fill=blue!50,draw=blue!50,line width=1mm]
(-1.46,2.0) -- (-1.46,2.732);
(-1.36,2.0) -- (-1.36,2.732);
\draw[dashed,blue!50] (-1.410,0.895) -- (-1.410,2.0); 
\draw[blue,fill=blue!100] (-1.672,0.633) circle (0.06cm); 
\path [fill=blue!50,draw=blue!50,line width=1mm]
(-1.722,2.0) -- (-1.722,2.762); 
(-1.622,2.0) -- (-1.622,2.762);
\draw[dashed,blue!50] (-1.672,0.633) -- (-1.672,2.0); 
\draw[blue,fill=blue!100] (-1.905,0.345) circle (0.06cm);  
\path [fill=blue!50,draw=blue!50,line width=1mm]
(-1.955,2.0) -- (-1.955,2.1); 
(-1.855,2.0) -- (-1.855,2.1);
\draw[dashed,blue!50] (-1.905,0.345) -- (-1.905,2.0); 
\draw[blue,fill=blue!100] (-2.109,0.036) circle (0.06cm);  
\path [fill=blue!50,draw=blue!50,line width=1mm]
(-2.159,2.0) -- (-2.159,2.814); 
(-2.059,2.0) -- (-2.059,2.814);
\draw[dashed,blue!50] (-2.109,0.036) -- (-2.109,2.0); 
\draw(-2.598,-1.500) -- (1.685,1.788);
\draw(1.685,1.788) -- (2.68,1.48);
\draw[black,fill=black!100] (-0.5,0.10992) circle (0.06cm);   
\coordinate[label={[below]\textbf{O}}] (a) at (-0.4,0.0);
\coordinate[label={[below]\textbf{S}}] (a) at (1.685,1.7);
\coordinate[label={[below]\textbf{A}}] (a) at (-2.2,-1.28);
\coordinate[label={[below]\textbf{B}}] (b) at (2.5,1.45);
\end{tikzpicture}}}\quad
  \subfigure[]{\scalebox{0.45}{\begin{tikzpicture}
\draw[line width=0.2mm] (0,0) circle (3cm); 
\draw [thick,blue,domain=63.719:174.988] plot ({1.049+3.661*cos(\x)}, {-1.817+3.661*sin(\x)});
\draw[dashed,blue!50] (-2.072,2.0) -- (1.735,2.0); 
\draw[blue,fill=blue!100] (1.685,1.788) circle (0.06cm); 
\path [fill=blue!50,draw=blue!50,line width=1mm]
(1.635,2.0) -- (1.635,2.1); 
(1.735,2.0) -- (1.735,2.1);
\draw[dashed,blue!50] (1.685,1.788) -- (1.685,2.0);
\draw[blue,fill=blue!100] (1.317,1.834) circle (0.06cm); 
\path [fill=blue!50,draw=blue!50,line width=1mm]
(1.317,2.0) -- (1.317,2.1);  
(1.267,2.0) -- (1.267,2.1);
\draw[dashed,blue!50] (1.317,1.834) -- (1.317,2.0);
\draw[blue,fill=blue!100] (0.947,1.843) circle (0.06cm); 
\path [fill=blue!50,draw=blue!50,line width=1mm]
(0.897,2.0) -- (0.897,2.329); 
(0.997,2.0) -- (0.997,2.329);
\draw[dashed,blue!50] (0.947,1.843) -- (0.947,2.0);
\draw[blue,fill=blue!100] (0.577,1.814) circle (0.06cm); 
\path [fill=blue!50,draw=blue!50,line width=1mm]
(0.527,2.0) -- (0.527,2.1); 
(0.627,2.0) -- (0.627,2.1);
\draw[dashed,blue!50] (0.577,1.814) -- (0.577,2.0);
\draw[blue,fill=blue!100] (0.213,1.747) circle (0.06cm); 
\path [fill=blue!50,draw=blue!50,line width=1mm]
(0.163,2.0) -- (0.163,2.329); 
(0.263,2.0) -- (0.263,2.329);
\draw[dashed,blue!50] (0.213,1.747) -- (0.213,2.0);
\draw[blue,fill=blue!100] (-0.143,1.645) circle (0.06cm); 
\path [fill=blue!50,draw=blue!50,line width=1mm]
(-0.193,2.0) -- (-0.193,2.329); 
(-0.093,2.0) -- (-0.093,2.329);
\draw[dashed,blue!50] (-0.143,1.643) -- (-0.143,2.0);
\draw[blue,fill=blue!100] (-0.487,1.506) circle (0.06cm); 
\path [fill=blue!50,draw=blue!50,line width=1mm]
(-0.537,2.0) -- (-0.537,2.329); 
(-0.437,2.0) -- (-0.437,2.329);
\draw[dashed,blue!50] (-0.487,1.506) -- (-0.487,2.0);
\draw[blue,fill=blue!100] (-0.815,1.334) circle (0.06cm);  
\path [fill=blue!50,draw=blue!50,line width=1mm]
(-0.865,2.0) -- (-0.865,2.1); 
(-0.765,2.0) -- (-0.765,2.1);
\draw[dashed,blue!50] (-0.815,1.334) -- (-0.815,2.0);
\draw[blue,fill=blue!100] (-1.124,1.130) circle (0.06cm);  
\path [fill=blue!50,draw=blue!50,line width=1mm]
(-1.174,2.0) -- (-1.174,2.1); 
(-1.074,2.0) -- (-1.074,2.1);
\draw[dashed,blue!50] (-1.124,1.130) -- (-1.124,2.0); 
\draw[red,fill=red!100] (-1.410,0.895) circle (0.06cm);  
\path [fill=blue!50,draw=blue!50,line width=1mm]
(-1.46,2.0) -- (-1.46,2.886); 
(-1.36,2.0) -- (-1.36,2.886);
\draw[dashed,blue!50] (-1.410,0.895) -- (-1.410,2.0); 
\draw[blue,fill=blue!100] (-1.672,0.633) circle (0.06cm); 
\path [fill=red!50,draw=red!50,line width=1mm]
(-1.722,2.0) -- (-1.722,3.0); 
(-1.622,2.0) -- (-1.622,3.0);
\draw[dashed,red!50] (-1.672,0.633) -- (-1.672,2.0); 
\draw[blue,fill=blue!100] (-1.905,0.345) circle (0.06cm);  
\path [fill=blue!50,draw=blue!50,line width=1mm]
(-1.955,2.0) -- (-1.955,2.241); 
(-1.855,2.0) -- (-1.855,2.241);
\draw[dashed,blue!50] (-1.905,0.345) -- (-1.905,2.0); 
\draw(-2.598,-1.500) -- (-1.410,0.895);
\draw(-1.410,0.895) -- (2.68,1.48);
\draw[black,fill=black!100] (-2.0,-0.29444) circle (0.06cm);   
\coordinate[label={[below]\textbf{O}}] (a) at (-1.7,-0.1);
\coordinate[label={[below]\textbf{S}}] (a) at (-1.3,0.85);
\coordinate[label={[below]\textbf{A}}] (a) at (-2.25,-1.28);
\coordinate[label={[below]\textbf{B}}] (b) at (2.5,1.45);
\end{tikzpicture}}}
  \caption{Pictorial depiction of the analyzed results (Table~\ref{table2}) for the ideal cases of energy and timing. The identified real scattering point corresponds to the maximum height of the bar (in red).}
  \label{fig10}
\end{figure}

We analyzed two data samples, but present here the results for only one of them. The bin size and the sample size were chosen to be $N$=65 and \textit{n}=6. Out of the six events, one was USE, and five were SSEs with Cardinalities 13, 40, 95, 12, and 27. The data were analyzed following the procedure explained in Sec.~\ref{sec:level4}. Table~2 presents the data with the lowest cardinalities 13 and 12; they are the typical examples consistent with all the empirical rules presented in Sec.~\ref{sec:level6}. As expected, the data value with the highest frequency (starred ``*'' in red) was for the real scattering point, \textit{i.e.}, the starred rows matched perfectly in Table~2 (left). However, in Table~2 (right), the highest frequency was obtained for a data value just adjacent to the true one; this was a reasonable observation because the two data values, 2474.67 and \textcolor{red}{2164.42*}, corresponded to nearby points on the circular arc (Fig.~\ref{fig10}(b)). The result was indeed found to be close to the real scattering point. \par

\subsection{\bf{Finite time resolution}}
\label{sec:level5_2}

Due to the finite timing resolution, annihilation points could not be located accurately. On the other hand, we found the TOF information to be redundant in performing the Virtual extrapolation, described earlier and also in \ref{appen1}. Hence, for checking, we performed the Virtual extrapolation for the simulated PET data (Sec.~\ref{sec:level3_1}) by placing the annihilation points arbitrarily anywhere on the LOR. For instance, the annihilation point could be at the unscattered photon detection point \textbf{A} in Fig.~\ref{fig6}. One such case of the data sample is presented below.
\par

\begin{figure}[htp]
  \centering
  \subfigure[]{\scalebox{0.45}{\begin{tikzpicture}
\draw[line width=0.2mm] (0,0) circle (3cm); 
\draw [thick,blue,domain=63.719:174.988] plot ({1.049+3.661*cos(\x)}, {-1.817+3.661*sin(\x)});
\draw[dashed,blue!50] (-2.109,2.0) -- (0.947,2.0);
\draw[blue,fill=blue!100] (0.947,1.843) circle (0.06cm); 
\path [fill=red!50,draw=red!50,line width=1mm]
(0.897,2.0) -- (0.897,3.0);                                  
(0.997,2.0) -- (0.997,3.0);
\draw[dashed,red!50] (0.947,1.843) -- (0.947,2.0);
\draw[red,fill=red!100] (0.577,1.814) circle (0.06cm); 
\path [fill=blue!50,draw=blue!50,line width=1mm]
(0.527,2.0) -- (0.527,2.1);                              
(0.627,2.0) -- (0.627,2.1);
\draw[dashed,blue!50] (0.577,1.814) -- (0.577,2.0);
\draw[blue,fill=blue!100] (0.213,1.747) circle (0.06cm); 
\path [fill=blue!50,draw=blue!50,line width=1mm]
(0.163,2.0) -- (0.163,2.928);                             
(0.263,2.0) -- (0.263,2.928);
\draw[dashed,blue!50] (0.213,1.747) -- (0.213,2.0);
\draw[blue,fill=blue!100] (-0.143,1.645) circle (0.06cm); 
\path [fill=blue!50,draw=blue!50,line width=1mm]
(-0.193,2.0) -- (-0.193,2.1);                                
(-0.093,2.0) -- (-0.093,2.1);
\draw[dashed,blue!50] (-0.143,1.643) -- (-0.143,2.0);
\draw[blue,fill=blue!100] (-0.487,1.506) circle (0.06cm); 
\path [fill=blue!50,draw=blue!50,line width=1mm]
(-0.537,2.0) -- (-0.537,2.1);                             
(-0.437,2.0) -- (-0.437,2.1);
\draw[dashed,blue!50] (-0.487,1.506) -- (-0.487,2.0);
\draw[blue,fill=blue!100] (-0.815,1.334) circle (0.06cm);  
\path [fill=blue!50,draw=blue!50,line width=1mm]
(-0.865,2.0) -- (-0.865,2.599);                               
(-0.765,2.0) -- (-0.765,2.599);
\draw[dashed,blue!50] (-0.815,1.334) -- (-0.815,2.0);
\draw[blue,fill=blue!100] (-1.124,1.130) circle (0.06cm);  
\path [fill=blue!50,draw=blue!50,line width=1mm]
(-1.174,2.0) -- (-1.174,2.1);
(-1.074,2.0) -- (-1.074,2.1);
\draw[dashed,blue!50] (-1.124,1.130) -- (-1.124,2.0); 
\draw[blue,fill=blue!100] (-1.410,0.895) circle (0.06cm);  
\path [fill=blue!50,draw=blue!50,line width=1mm]
(-1.46,2.0) -- (-1.46,2.896);
(-1.36,2.0) -- (-1.36,2.896);
\draw[dashed,blue!50] (-1.410,0.895) -- (-1.410,2.0); 
\draw[blue,fill=blue!100] (-1.672,0.633) circle (0.06cm); 
\path [fill=blue!50,draw=blue!50,line width=1mm]
(-1.722,2.0) -- (-1.722,2.597);
(-1.622,2.0) -- (-1.622,2.597);
\draw[dashed,blue!50] (-1.672,0.633) -- (-1.672,2.0); 
\draw[blue,fill=blue!100] (-1.905,0.345) circle (0.06cm);  
\path [fill=blue!50,draw=blue!50,line width=1mm]
(-1.955,2.0) -- (-1.955,2.925);
(-1.855,2.0) -- (-1.855,2.925);
\draw[dashed,blue!50] (-1.905,0.345) -- (-1.905,2.0); 
\draw[blue,fill=blue!100] (-2.109,0.036) circle (0.06cm);  
\path [fill=blue!50,draw=blue!50,line width=1mm]
(-2.159,2.0) -- (-2.159,2.597);
(-2.059,2.0) -- (-2.059,2.597);
\draw[dashed,blue!50] (-2.109,0.036) -- (-2.109,2.0); 
\draw(-2.598,-1.500) -- (0.577,1.814);
\draw(0.577,1.814) -- (2.68,1.48);
\draw[black,fill=black!100] (-1.0,0.168312) circle (0.06cm);
\coordinate[label={[below]\textbf{O}}] (a) at (-0.7,0.168312);
\coordinate[label={[below]\textbf{S}}] (a) at (0.577,1.7);
\coordinate[label={[below]\textbf{A}}] (a) at (-2.25,-1.28);
\coordinate[label={[below]\textbf{B}}] (b) at (2.5,1.45);
\end{tikzpicture}}}\quad
  \subfigure[]{\scalebox{0.45}{\begin{tikzpicture}
\draw[line width=0.2mm] (0,0) circle (3cm); 
\draw [thick,blue,domain=63.719:174.988] plot ({1.049+3.661*cos(\x)}, {-1.817+3.661*sin(\x)});
\draw[dashed,blue!50] (0.577,2.0) -- (-2.109,2.0);
\draw[blue,fill=blue!100] (0.577,1.814) circle (0.06cm); 
\path [fill=blue!50,draw=blue!50,line width=1mm]
(0.527,2.0) -- (0.527,2.825);                              
(0.627,2.0) -- (0.627,2.825);
\draw[dashed,blue!50] (0.577,1.814) -- (0.577,2.0);
\draw[blue,fill=blue!100] (0.213,1.747) circle (0.06cm); 
\path [fill=blue!50,draw=blue!50,line width=1mm]
(0.163,2.0) -- (0.163,2.905);                                
(0.263,2.0) -- (0.263,2.905);
\draw[dashed,blue!50] (0.213,1.747) -- (0.213,2.0);
\draw[blue,fill=blue!100] (-0.143,1.645) circle (0.06cm); 
\path [fill=blue!50,draw=blue!50,line width=1mm]
(-0.193,2.0) -- (-0.193,2.745);                               
(-0.093,2.0) -- (-0.093,2.745);
\draw[dashed,blue!50] (-0.143,1.643) -- (-0.143,2.0);
\draw[blue,fill=blue!100] (-0.487,1.506) circle (0.06cm); 
\path [fill=blue!50,draw=blue!50,line width=1mm]
(-0.537,2.0) -- (-0.537,2.969);                               
(-0.437,2.0) -- (-0.437,2.969);
\draw[dashed,blue!50] (-0.487,1.506) -- (-0.487,2.0);
\draw[blue,fill=blue!100] (-0.815,1.334) circle (0.06cm);  
\path [fill=blue!50,draw=blue!50,line width=1mm]
(-0.865,2.0) -- (-0.865,2.1);                               
(-0.765,2.0) -- (-0.765,2.1);
\draw[dashed,blue!50] (-0.815,1.334) -- (-0.815,2.0);
\draw[blue,fill=blue!100] (-1.124,1.130) circle (0.06cm);  
\path [fill=blue!50,draw=blue!50,line width=1mm]
(-1.174,2.0) -- (-1.174,2.1);
(-1.074,2.0) -- (-1.074,2.1);
\draw[dashed,blue!50] (-1.124,1.130) -- (-1.124,2.0); 
\draw[blue,fill=blue!100] (-1.410,0.895) circle (0.06cm);  
\path [fill=blue!50,draw=blue!50,line width=1mm]
(-1.46,2.0) -- (-1.46,2.864);
(-1.36,2.0) -- (-1.36,2.864);
\draw[dashed,blue!50] (-1.410,0.895) -- (-1.410,2.0); 
\draw[blue,fill=blue!100] (-1.672,0.633) circle (0.06cm); 
\path [fill=red!50,draw=red!50,line width=1mm]
(-1.722,2.0) -- (-1.722,3.0);
(-1.622,2.0) -- (-1.622,3.0);
\draw[dashed,red!50] (-1.672,0.633) -- (-1.672,2.0); 
\draw[red,fill=red!100] (-1.905,0.345) circle (0.06cm);  
\path [fill=blue!50,draw=blue!50,line width=1mm]
(-1.955,2.0) -- (-1.955,2.1);
(-1.855,2.0) -- (-1.855,2.1);
\draw[dashed,blue!50] (-1.905,0.345) -- (-1.905,2.0); 
\draw[blue,fill=blue!100] (-2.109,0.036) circle (0.06cm);  
\path [fill=blue!50,draw=blue!50,line width=1mm]
(-2.159,2.0) -- (-2.159,2.745);
(-2.059,2.0) -- (-2.059,2.745);
\draw[dashed,blue!50] (-2.109,0.036) -- (-2.109,2.0); 
\draw(-2.598,-1.500) -- (-1.905,0.345);
\draw(-1.905,0.345) -- (2.68,1.48);
\draw[black,fill=black!100] (-2.2,-0.44029) circle (0.06cm);
\coordinate[label={[below]\textbf{O}}] (a) at (-1.9,-0.38);
\coordinate[label={[below]\textbf{S}}] (a) at (-1.7,0.3);
\coordinate[label={[below]\textbf{A}}] (a) at (-2.25,-1.28);
\coordinate[label={[below]\textbf{B}}] (b) at (2.5,1.45);
\end{tikzpicture}}}
  \subfigure[]{\scalebox{0.45}{\begin{tikzpicture}
\draw[line width=0.2mm] (0,0) circle (3cm); 
\draw [thick,blue,domain=63.719:174.988] plot ({1.049+3.661*cos(\x)}, {-1.817+3.661*sin(\x)});
\draw[dashed,blue!50] (-0.143,2.0) -- (-1.905,2.0);
\draw[blue,fill=blue!100] (-0.143,1.645) circle (0.06cm); 
\path [fill=blue!50,draw=blue!50,line width=1mm]
(-0.193,2.0) -- (-0.193,2.197);                             
(-0.093,2.0) -- (-0.093,2.197);
\draw[dashed,blue!50] (-0.143,1.643) -- (-0.143,2.0);
\draw[blue,fill=blue!100] (-0.487,1.506) circle (0.06cm); 
\path [fill=blue!50,draw=blue!50,line width=1mm]
(-0.537,2.0) -- (-0.537,2.888);                         
(-0.437,2.0) -- (-0.437,2.888);
\draw[dashed,blue!50] (-0.487,1.506) -- (-0.487,2.0);
\draw[blue,fill=blue!100] (-0.815,1.334) circle (0.06cm);  
\path [fill=blue!50,draw=blue!50,line width=1mm]
(-0.865,2.0) -- (-0.865,2.1);                             
(-0.765,2.0) -- (-0.765,2.1);
\draw[dashed,blue!50] (-0.815,1.334) -- (-0.815,2.0);
\draw[blue,fill=blue!100] (-1.124,1.130) circle (0.06cm);  
\path [fill=blue!50,draw=blue!50,line width=1mm]
(-1.174,2.0) -- (-1.174,2.598);
(-1.074,2.0) -- (-1.074,2.598);
\draw[dashed,blue!50] (-1.124,1.130) -- (-1.124,2.0); 
\draw[blue,fill=blue!100] (-1.410,0.895) circle (0.06cm);  
\path [fill=blue!50,draw=blue!50,line width=1mm]
(-1.46,2.0) -- (-1.46,2.955);
(-1.36,2.0) -- (-1.36,2.955);
\draw[dashed,blue!50] (-1.410,0.895) -- (-1.410,2.0); 
\draw[red,fill=red!100] (-1.672,0.633) circle (0.06cm); 
\path [fill=blue!50,draw=blue!50,line width=1mm]
(-1.722,2.0) -- (-1.722,2.888);
(-1.622,2.0) -- (-1.622,2.888);
\draw[dashed,blue!50] (-1.672,0.633) -- (-1.672,2.0); 
\draw[blue,fill=blue!100] (-1.905,0.345) circle (0.06cm);  
\path [fill=red!50,draw=red!50,line width=1mm]
(-1.955,2.0) -- (-1.955,3.0);
(-1.855,2.0) -- (-1.855,3.0);
\draw[dashed,red!50] (-1.905,0.345) -- (-1.905,2.0); 
\draw(-2.598,-1.500) -- (-1.672,0.633);
\draw(-1.672,0.633) -- (2.68,1.48);
\draw[black,fill=black!100] (-2.0,-0.122) circle (0.06cm);
\coordinate[label={[below]\textbf{O}}] (a) at (-1.7,-0.08);
\coordinate[label={[below]\textbf{S}}] (a) at (-1.5,0.52);
\coordinate[label={[below]\textbf{A}}] (a) at (-2.25,-1.28);
\coordinate[label={[below]\textbf{B}}] (b) at (2.5,1.45);
\end{tikzpicture}}}\quad
  \subfigure[]{\scalebox{0.45}{\begin{tikzpicture}
\draw[line width=0.2mm] (0,0) circle (3cm); 
\draw [thick,blue,domain=63.719:174.988] plot ({1.049+3.661*cos(\x)}, {-1.817+3.661*sin(\x)});
\draw[dashed,blue!50] (-0.143,2.0) -- (-1.905,2.0);
\draw[blue,fill=blue!100] (-0.143,1.645) circle (0.06cm); 
\path [fill=red!50,draw=red!50,line width=1mm]
(-0.193,2.0) -- (-0.193,3.0);                                   
(-0.093,2.0) -- (-0.093,3.0);
\draw[dashed,red!50] (-0.143,1.643) -- (-0.143,2.0);
\draw[red,fill=red!100] (-0.487,1.506) circle (0.06cm); 
\path [fill=blue!50,draw=blue!50,line width=1mm]
(-0.537,2.0) -- (-0.537,2.1);                              
(-0.437,2.0) -- (-0.437,2.1);
\draw[dashed,blue!50] (-0.487,1.506) -- (-0.487,2.0);
\draw[blue,fill=blue!100] (-0.815,1.334) circle (0.06cm);  
\path [fill=blue!50,draw=blue!50,line width=1mm]
(-0.865,2.0) -- (-0.865,2.977);                              
(-0.765,2.0) -- (-0.765,2.977);
\draw[dashed,blue!50] (-0.815,1.334) -- (-0.815,2.0);
\draw[blue,fill=blue!100] (-1.124,1.130) circle (0.06cm);  
\path [fill=blue!50,draw=blue!50,line width=1mm]
(-1.174,2.0) -- (-1.174,2.752);
(-1.074,2.0) -- (-1.074,2.752);
\draw[dashed,blue!50] (-1.124,1.130) -- (-1.124,2.0); 
\draw[blue,fill=blue!100] (-1.410,0.895) circle (0.06cm);  
\path [fill=blue!50,draw=blue!50,line width=1mm]
(-1.46,2.0) -- (-1.46,2.1);
(-1.36,2.0) -- (-1.36,2.1);
\draw[dashed,blue!50] (-1.410,0.895) -- (-1.410,2.0); 
\draw[blue,fill=blue!100] (-1.672,0.633) circle (0.06cm); 
\path [fill=blue!50,draw=blue!50,line width=1mm]
(-1.722,2.0) -- (-1.722,2.752);
(-1.622,2.0) -- (-1.622,2.752);
\draw[dashed,blue!50] (-1.672,0.633) -- (-1.672,2.0); 
\draw[blue,fill=blue!100] (-1.905,0.345) circle (0.06cm);  
\path [fill=blue!50,draw=blue!50,line width=1mm]
(-1.955,2.0) -- (-1.955,2.752);
(-1.855,2.0) -- (-1.855,2.752);
\draw[dashed,blue!50] (-1.905,0.345) -- (-1.905,2.0); 
\draw(-2.598,-1.500) -- (-0.487,1.506);
\draw(-0.487,1.506) -- (2.68,1.48);
\draw[black,fill=black!100] (-1.5,0.063) circle (0.06cm); 
\coordinate[label={[below]\textbf{O}}] (a) at (-1.2,0.2);
\coordinate[label={[below]\textbf{S}}] (a) at (-0.37,1.3);
\coordinate[label={[below]\textbf{A}}] (a) at (-2.25,-1.28);
\coordinate[label={[below]\textbf{B}}] (b) at (2.5,1.45);
\end{tikzpicture}}}
  \caption{Pictorial depiction of the convergence results for the case of finite timing resolution.}
  \label{fig11}
\end{figure}
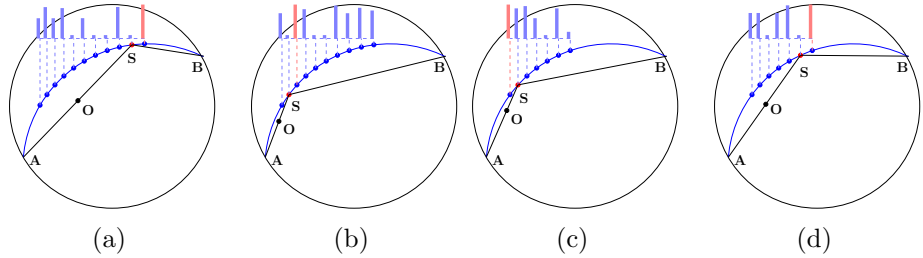

The sample size (\textit{n}) was eight containing two USEs and six SSEs with cardinalities 104, 174, 11, 10, 7, and 7. The total number of bins (\textit{N}) was 107. Figure~\ref{fig11} presents the results pictorially for the last four events with low cardinalities 11, 10, 7, and 7. In all these events, we indeed got the maximum frequency -- but not distinctly very high -- for the data values very close to the real scattering points.

\subsection{\bf{Finite energy resolution}}
\label{sec:level5_3}

\begin{table*}[htb!]
\centering
\begin{minipage}[b]{0.45\linewidth}
\centering
\begin{tabular}{||c | c | c||}
\hline
No. & z (mm) & Frequency  \\ [0.5ex] 
\hline\hline
1	&	1753.95	&	\textcolor{red}{$64784766^{*}$}	\\	\hline
2	&	\textcolor{red}{$1576.04^{*}$}	&	64363043	\\	\hline
3	&	3407.86	&	62786354	\\	\hline
4	&	1807.26	&	\textcolor{green}{$64784766^{*}$}	\\	\hline
5	&	873.386	&	59256539	\\	\hline
6	&	708.819	&	64047177	\\	\hline
7	&	1994.78	&	64387783	\\	\hline
\end{tabular}
\end{minipage}
\begin{minipage}[b]{0.45\linewidth}
\centering
\begin{tabular}{||c | c | c||} 
\hline
No. & z (mm) & Frequency  \\ [0.5ex] 
\hline\hline
1	&	2249.57	&	40648295	\\	\hline
2	&	\textcolor{red}{$1134.35^{*}$}	&	40465536	\\	\hline
3	&	1760.18	&	\textcolor{red}{${41113288}^{*}$}	\\	\hline
4	&	2221.58	&	40648295	\\	\hline
5	&	2001.11	&	40607369	\\	\hline
6	&	2275.41	&	40648295	\\	\hline
7	&	1797.53	&	\textcolor{green}{$41113288^{*}$}	\\	\hline
8	&	874.281	&	37729165	\\	\hline
9	&	1311.37	&	40474414	\\	\hline
10	&	4424	&	40488069	\\	\hline
11	&	1236.12	&	40474414	\\	\hline
\end{tabular}
\end{minipage}

\caption{Analyzed results for the data samples with cardinality 7 (left) and cardinality 11 (right), for the case of finite energy resolution of detectors; see details in the text.}
\label{table3}
\end{table*}

We implemented an energy resolution of 10\% FWHM at 511 keV on the simulated PET data by applying Gaussian blurring on the scattered photon energies. We also considered an energy dependence of $\frac{1}{\sqrt{E}}$ for the photon of energy $E$~\cite{conti2012reconstruction}. As a result, we obtained a blurring in the circular arc for the scattering points (Sec.~\ref{sec:level3_3_1}), creating two equidistant arcs from the original arc (Fig.~\ref{fig7}). We performed Virtual extrapolation for all possible scattering points on the displaced circular loci.\par

\begin{figure}[htp]
  \centering
  \subfigure[]{\scalebox{0.43}{\begin{tikzpicture}
\draw[line width=0.2mm] (0,0) circle (3cm); 
\draw [thick,blue,domain=63.719:174.988] plot ({1.049+3.661*cos(\x)}, {-1.817+3.661*sin(\x)});
\draw[dashed,blue!50] (-0.143,2.0) -- (-1.905,2.0);
\draw[blue,fill=blue!100] (-0.143,1.645) circle (0.06cm); 
\path [fill=red!50,draw=red!50,line width=1mm]
(-0.193,2.0) -- (-0.193,3.0);  
(-0.093,2.0) -- (-0.093,3.0);
\draw[dashed,red!50] (-0.143,1.643) -- (-0.143,2.0);
\draw[red,fill=red!100] (-0.487,1.506) circle (0.06cm); 
\path [fill=blue!50,draw=blue!50,line width=1mm]
(-0.537,2.0) -- (-0.537,2.931); 
(-0.437,2.0) -- (-0.437,2.931);
\draw[dashed,blue!50] (-0.487,1.506) -- (-0.487,2.0);
\draw[blue,fill=blue!100] (-0.815,1.334) circle (0.06cm);  
\path [fill=blue!50,draw=blue!50,line width=1mm]
(-0.865,2.0) -- (-0.865,2.674); 
(-0.765,2.0) -- (-0.765,2.674);
\draw[dashed,blue!50] (-0.815,1.334) -- (-0.815,2.0);
\draw[blue,fill=blue!100] (-1.124,1.130) circle (0.06cm);  
\path [fill=green!50,draw=green!50,line width=1mm]
(-1.174,2.0) -- (-1.174,3.0);   
(-1.074,2.0) -- (-1.074,3.0);
\draw[dashed,green!50] (-1.124,1.130) -- (-1.124,2.0); 
\draw[blue,fill=blue!100] (-1.410,0.895) circle (0.06cm);  
\path [fill=blue!50,draw=blue!50,line width=1mm]
(-1.46,2.0) -- (-1.46,2.1);  
(-1.36,2.0) -- (-1.36,2.1);
\draw[dashed,blue!50] (-1.410,0.895) -- (-1.410,2.0); 
\draw[blue,fill=blue!100] (-1.672,0.633) circle (0.06cm); 
\path [fill=blue!50,draw=blue!50,line width=1mm]
(-1.722,2.0) -- (-1.722,2.880); 
(-1.622,2.0) -- (-1.622,2.880);
\draw[dashed,blue!50] (-1.672,0.633) -- (-1.672,2.0); 
\draw[blue,fill=blue!100] (-1.905,0.345) circle (0.06cm);  
\path [fill=blue!50,draw=blue!50,line width=1mm]
(-1.955,2.0) -- (-1.955,2.935); 
(-1.855,2.0) -- (-1.855,2.935);
\draw[dashed,blue!50] (-1.905,0.345) -- (-1.905,2.0); 
\draw(-2.598,-1.500) -- (-0.487,1.506);
\draw(-0.487,1.506) -- (2.68,1.48);
\draw[black,fill=black!100] (-1.5,0.063) circle (0.06cm);   
\coordinate[label={[below]\textbf{O}}] (a) at (-1.2,0.2);
\coordinate[label={[below]\textbf{S}}] (a) at (-0.37,1.3);
\coordinate[label={[below]\textbf{A}}] (a) at (-2.25,-1.28);
\coordinate[label={[below]\textbf{B}}] (b) at (2.5,1.45);
\end{tikzpicture}}}\quad
  \subfigure[]{\scalebox{0.43}{\begin{tikzpicture}
\draw[line width=0.2mm] (0,0) circle (3cm); 
\draw [thick,blue,domain=63.719:174.988] plot ({1.049+3.661*cos(\x)}, {-1.817+3.661*sin(\x)});
\draw[dashed,blue!50] (-0.143,2.0) -- (-1.905,2.0); 
\draw[blue,fill=blue!100] (1.317,1.834) circle (0.06cm); 
\path [fill=blue!50,draw=blue!50,line width=1mm]
(1.317,2.0) -- (1.317,2.876);  
(1.267,2.0) -- (1.267,2.876);
\draw[dashed,blue!50] (1.317,1.834) -- (1.317,2.0);
\draw[red,fill=red!100] (0.947,1.843) circle (0.06cm); 
\path [fill=blue!50,draw=blue!50,line width=1mm]
(0.897,2.0) -- (0.897,2.876);  
(0.997,2.0) -- (0.997,2.876);
\draw[dashed,blue!50] (0.947,1.843) -- (0.947,2.0);
\draw[blue,fill=blue!100] (0.577,1.814) circle (0.06cm); 
\path [fill=red!50,draw=red!50,line width=1mm]
(0.527,2.0) -- (0.527,3.0);   
(0.627,2.0) -- (0.627,3.0);
\draw[dashed,red!50] (0.577,1.814) -- (0.577,2.0);
\draw[blue,fill=blue!100] (0.213,1.747) circle (0.06cm); 
\path [fill=blue!50,draw=blue!50,line width=1mm]
(0.163,2.0) -- (0.163,2.876);  
(0.263,2.0) -- (0.263,2.876);
\draw[dashed,blue!50] (0.213,1.747) -- (0.213,2.0);
\draw[blue,fill=blue!100] (-0.143,1.645) circle (0.06cm); 
\path [fill=blue!50,draw=blue!50,line width=1mm]
(-0.193,2.0) -- (-0.193,2.865); 
(-0.093,2.0) -- (-0.093,2.865);
\draw[dashed,blue!50] (-0.143,1.643) -- (-0.143,2.0);
\draw[blue,fill=blue!100] (-0.487,1.506) circle (0.06cm); 
\path [fill=blue!50,draw=blue!50,line width=1mm]
(-0.537,2.0) -- (-0.537,2.876); 
(-0.437,2.0) -- (-0.437,2.876);
\draw[dashed,blue!50] (-0.487,1.506) -- (-0.487,2.0);
\draw[blue,fill=blue!100] (-0.815,1.334) circle (0.06cm);  
\path [fill=green!50,draw=green!50,line width=1mm]
(-0.865,2.0) -- (-0.865,3.0); 
(-0.765,2.0) -- (-0.765,3.0);
\draw[dashed,green!50] (-0.815,1.334) -- (-0.815,2.0);
\draw[blue,fill=blue!100] (-1.124,1.130) circle (0.06cm);  
\path [fill=blue!50,draw=blue!50,line width=1mm]
(-1.174,2.0) -- (-1.174,2.1); 
(-1.074,2.0) -- (-1.074,2.1);
\draw[dashed,blue!50] (-1.124,1.130) -- (-1.124,2.0); 
\draw[blue,fill=blue!100] (-1.410,0.895) circle (0.06cm);  
\path [fill=blue!50,draw=blue!50,line width=1mm]
(-1.46,2.0) -- (-1.46,2.830); 
(-1.36,2.0) -- (-1.36,2.830);
\draw[dashed,blue!50] (-1.410,0.895) -- (-1.410,2.0); 
\draw[blue,fill=blue!100] (-1.672,0.633) circle (0.06cm); 
\path [fill=blue!50,draw=blue!50,line width=1mm]
(-1.722,2.0) -- (-1.722,2.834); 
(-1.622,2.0) -- (-1.622,2.834);
\draw[dashed,blue!50] (-1.672,0.633) -- (-1.672,2.0); 
\draw[blue,fill=blue!100] (-1.905,0.345) circle (0.06cm);  
\path [fill=blue!50,draw=blue!50,line width=1mm]
(-1.955,2.0) -- (-1.955,2.830); 
(-1.855,2.0) -- (-1.855,2.830);
\draw[dashed,blue!50] (-1.905,0.345) -- (-1.905,2.0); 
\draw(-2.598,-1.500) -- (0.947,1.843); 
\draw(0.947,1.843) -- (2.68,1.48);
\draw[black,fill=black!100] (-0.7,0.2899) circle (0.06cm);   
\coordinate[label={[below]\textbf{O}}] (a) at (-0.5,0.29);
\coordinate[label={[below]\textbf{S}}] (a) at (0.947,1.6);
\coordinate[label={[below]\textbf{A}}] (a) at (-2.25,-1.28);
\coordinate[label={[below]\textbf{B}}] (b) at (2.5,1.45);
\end{tikzpicture}}}
  \subfigure[]{\scalebox{0.43}{\begin{tikzpicture}
\draw[line width=0.2mm] (0,0) circle (3cm); 
\draw [thick,blue,domain=63.719:174.988] plot ({1.049+3.661*cos(\x)}, {-1.817+3.661*sin(\x)});
\draw[dashed,blue!50] (-0.143,2.0) -- (-1.905,2.0); 
\draw[blue,fill=blue!100] (0.947,1.843) circle (0.06cm); 
\path [fill=red!50,draw=red!50,line width=1mm]
(0.897,2.0) -- (0.897,3.0);   
(0.997,2.0) -- (0.997,3.0);
\draw[dashed,red!50] (0.947,1.843) -- (0.947,2.0);
\draw[red,fill=red!100] (0.577,1.814) circle (0.06cm); 
\path [fill=blue!50,draw=blue!50,line width=1mm]
(0.527,2.0) -- (0.527,2.908);  
(0.627,2.0) -- (0.627,2.908);
\draw[dashed,blue!50] (0.577,1.814) -- (0.577,2.0);
\draw[blue,fill=blue!100] (0.213,1.747) circle (0.06cm); 
\path [fill=blue!50,draw=blue!50,line width=1mm]
(0.163,2.0) -- (0.163,2.634);  
(0.263,2.0) -- (0.263,2.634);
\draw[dashed,blue!50] (0.213,1.747) -- (0.213,2.0);
\draw[blue,fill=blue!100] (-0.143,1.645) circle (0.06cm); 
\path [fill=blue!50,draw=blue!50,line width=1mm]
(-0.193,2.0) -- (-0.193,2.1); 
(-0.093,2.0) -- (-0.093,2.1);
\draw[dashed,red!50] (-0.143,1.643) -- (-0.143,2.0);
\draw[blue,fill=blue!100] (-0.487,1.506) circle (0.06cm); 
\path [fill=blue!50,draw=blue!50,line width=1mm]
(-0.537,2.0) -- (-0.537,2.984); 
(-0.437,2.0) -- (-0.437,2.984);
\draw[dashed,blue!50] (-0.487,1.506) -- (-0.487,2.0);
\draw[blue,fill=blue!100] (-0.815,1.334) circle (0.06cm);  
\path [fill=blue!50,draw=blue!50,line width=1mm]
(-0.865,2.0) -- (-0.865,2.984); 
(-0.765,2.0) -- (-0.765,2.984);
\draw[dashed,blue!50] (-0.815,1.334) -- (-0.815,2.0);
\draw[blue,fill=blue!100] (-1.124,1.130) circle (0.06cm);  
\path [fill=blue!50,draw=blue!50,line width=1mm]
(-1.174,2.0) -- (-1.174,2.437); 
(-1.074,2.0) -- (-1.074,2.437);
\draw[dashed,blue!50] (-1.124,1.130) -- (-1.124,2.0); 
\draw[blue,fill=blue!100] (-1.410,0.895) circle (0.06cm);  
\path [fill=blue!50,draw=blue!50,line width=1mm]
(-1.46,2.0) -- (-1.46,2.437);   
(-1.36,2.0) -- (-1.36,2.437);
\draw[dashed,blue!50] (-1.410,0.895) -- (-1.410,2.0); 
\draw[blue,fill=blue!100] (-1.672,0.633) circle (0.06cm); 
\path [fill=blue!50,draw=blue!50,line width=1mm]
(-1.722,2.0) -- (-1.722,2.437);  
(-1.622,2.0) -- (-1.622,2.437);
\draw[dashed,blue!50] (-1.672,0.633) -- (-1.672,2.0); 
\draw[blue,fill=blue!100] (-1.905,0.345) circle (0.06cm);  
\path [fill=blue!50,draw=blue!50,line width=1mm]
(-1.955,2.0) -- (-1.955,2.856); 
(-1.855,2.0) -- (-1.855,2.856);
\draw[dashed,blue!50] (-1.905,0.345) -- (-1.905,2.0); 
\draw(-2.598,-1.500) -- (0.577,1.814);
\draw(0.577,1.814) -- (2.68,1.48);
\draw[black,fill=black!100] (-1.0,0.167612) circle (0.06cm);   
\coordinate[label={[below]\textbf{O}}] (a) at (-0.8,0.14);
\coordinate[label={[below]\textbf{S}}] (a) at (0.577,1.6);
\coordinate[label={[below]\textbf{A}}] (a) at (-2.25,-1.28);
\coordinate[label={[below]\textbf{B}}] (b) at (2.5,1.45);
\end{tikzpicture}}}\quad
  \subfigure[]{\scalebox{0.43}{\begin{tikzpicture}
\draw[line width=0.2mm] (0,0) circle (3cm); 
\draw [thick,blue,domain=63.719:174.988] plot ({1.049+3.661*cos(\x)}, {-1.817+3.661*sin(\x)});
\draw[dashed,blue!50] (-0.143,2.0) -- (-1.905,2.0); 
\draw[blue,fill=blue!100] (-0.143,1.645) circle (0.06cm); 
\path [fill=blue!50,draw=blue!50,line width=1mm]
(-0.193,2.0) -- (-0.193,2.649); 
(-0.093,2.0) -- (-0.093,2.649);
\draw[dashed,blue!50] (-0.143,1.643) -- (-0.143,2.0);
\draw[red,fill=red!100] (-0.487,1.506) circle (0.06cm); 
\path [fill=red!50,draw=red!50,line width=1mm]
(-0.537,2.0) -- (-0.537,3.0); 
(-0.437,2.0) -- (-0.437,3.0);
\draw[dashed,red!50] (-0.487,1.506) -- (-0.487,2.0);
\draw[blue,fill=blue!100] (-0.815,1.334) circle (0.06cm);  
\path [fill=blue!50,draw=blue!50,line width=1mm]
(-0.865,2.0) -- (-0.865,2.657); 
(-0.765,2.0) -- (-0.765,2.657);
\draw[dashed,blue!50] (-0.815,1.334) -- (-0.815,2.0);
\draw[blue,fill=blue!100] (-1.124,1.130) circle (0.06cm);  
\path [fill=blue!50,draw=blue!50,line width=1mm]
(-1.174,2.0) -- (-1.174,2.633); 
(-1.074,2.0) -- (-1.074,2.633);
\draw[dashed,blue!50] (-1.124,1.130) -- (-1.124,2.0); 
\draw[blue,fill=blue!100] (-1.410,0.895) circle (0.06cm);  
\path [fill=blue!50,draw=blue!50,line width=1mm]
(-1.46,2.0) -- (-1.46,2.8); 
(-1.36,2.0) -- (-1.36,2.8);
\draw[dashed,blue!50] (-1.410,0.895) -- (-1.410,2.0); 
\draw[blue,fill=blue!100] (-1.672,0.633) circle (0.06cm); 
\path [fill=blue!50,draw=blue!50,line width=1mm]
(-1.722,2.0) -- (-1.722,2.1); 
(-1.622,2.0) -- (-1.622,2.1);
\draw[dashed,blue!50] (-1.672,0.633) -- (-1.672,2.0); 
\draw[blue,fill=blue!100] (-1.905,0.345) circle (0.06cm);  
\path [fill=blue!50,draw=blue!50,line width=1mm]
(-1.955,2.0) -- (-1.955,2.716); 
(-1.855,2.0) -- (-1.855,2.716);
\draw[dashed,blue!50] (-1.905,0.345) -- (-1.905,2.0); 
\draw(-2.598,-1.500) -- (-0.487,1.506);
\draw(-0.487,1.506) -- (2.68,1.48);
\draw[black,fill=black!100] (-1.5,0.063) circle (0.06cm);   
\coordinate[label={[below]\textbf{O}}] (a) at (-1.2,0.2);
\coordinate[label={[below]\textbf{S}}] (a) at (-0.37,1.3);
\coordinate[label={[below]\textbf{A}}] (a) at (-2.25,-1.28);
\coordinate[label={[below]\textbf{B}}] (b) at (2.5,1.45);
\end{tikzpicture}}}
  \subfigure[]{\scalebox{0.43}{\begin{tikzpicture}
\draw[line width=0.2mm] (0,0) circle (3cm); 
\draw [thick,blue,domain=63.719:174.988] plot ({1.049+3.661*cos(\x)}, {-1.817+3.661*sin(\x)});
\draw[dashed,blue!50] (-0.143,2.0) -- (-1.905,2.0); 
\draw[blue,fill=blue!100] (-0.143,1.645) circle (0.06cm); 
\path [fill=blue!50,draw=blue!50,line width=1mm]
(-0.193,2.0) -- (-0.193,2.499);  
(-0.093,2.0) -- (-0.093,2.499);
\draw[dashed,blue!50] (-0.143,1.643) -- (-0.143,2.0);
\draw[red,fill=red!100] (-0.487,1.506) circle (0.06cm); 
\path [fill=blue!50,draw=blue!50,line width=1mm]
(-0.537,2.0) -- (-0.537,2.635);  
(-0.437,2.0) -- (-0.437,2.635);
\draw[dashed,blue!50] (-0.487,1.506) -- (-0.487,2.0);
\draw[blue,fill=blue!100] (-0.815,1.334) circle (0.06cm);  
\path [fill=red!50,draw=red!50,line width=1mm]
(-0.865,2.0) -- (-0.865,3.0);    
(-0.765,2.0) -- (-0.765,3.0);
\draw[dashed,red!50] (-0.815,1.334) -- (-0.815,2.0);
\draw[blue,fill=blue!100] (-1.124,1.130) circle (0.06cm);  
\path [fill=blue!50,draw=blue!50,line width=1mm]
(-1.174,2.0) -- (-1.174,2.947);  
(-1.074,2.0) -- (-1.074,2.947);
\draw[dashed,blue!50] (-1.124,1.130) -- (-1.124,2.0); 
\draw[blue,fill=blue!100] (-1.410,0.895) circle (0.06cm);  
\path [fill=blue!50,draw=blue!50,line width=1mm]
(-1.46,2.0) -- (-1.46,2.917);    
(-1.36,2.0) -- (-1.36,2.917);
\draw[dashed,blue!50] (-1.410,0.895) -- (-1.410,2.0); 
\draw[blue,fill=blue!100] (-1.672,0.633) circle (0.06cm); 
\path [fill=blue!50,draw=blue!50,line width=1mm]
(-1.722,2.0) -- (-1.722,2.1); 
(-1.622,2.0) -- (-1.622,2.1);
\draw[dashed,blue!50] (-1.672,0.633) -- (-1.672,2.0); 
\draw[blue,fill=blue!100] (-1.905,0.345) circle (0.06cm);  
\path [fill=blue!50,draw=blue!50,line width=1mm]
(-1.955,2.0) -- (-1.955,2.661); 
(-1.855,2.0) -- (-1.855,2.661);
\draw[dashed,blue!50] (-1.905,0.345) -- (-1.905,2.0); 
\draw(-2.598,-1.500) -- (-0.487,1.506);
\draw(-0.487,1.506) -- (2.68,1.48);
\draw[black,fill=black!100] (-1.5,0.063) circle (0.06cm);   
\coordinate[label={[below]\textbf{O}}] (a) at (-1.2,0.2);
\coordinate[label={[below]\textbf{S}}] (a) at (-0.37,1.3);
\coordinate[label={[below]\textbf{A}}] (a) at (-2.25,-1.28);
\coordinate[label={[below]\textbf{B}}] (b) at (2.5,1.45);
\end{tikzpicture}}}\quad
  \caption{Pictorial depiction of the convergence results for the case of finite energy resolution.}
  \label{fig12}
\end{figure}
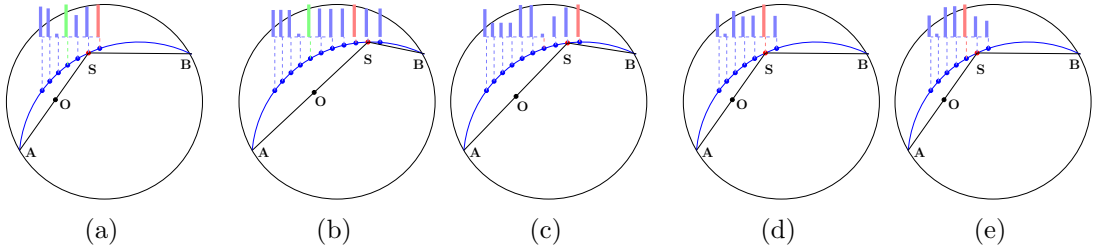

Here we present the results for one data sample. The sample size (\textit{n}) was nine containing two USEs, and seven SSEs with cardinalities 7, 97, 174, 11, 10, 7, and 7. The total number of bins (\textit{N}) was 145. The data values for the real scattering points corresponding to the highest frequency were recognized for five events with low cardinalities 7, 11, 10, 7, and 7. Figure~\ref{fig12} presents the results pictorially for these events. In all these events, we got the highest frequencies at the expected scattering points, consistent with our claim of the Second symmetric circle. However, in both parts (a) and (b) of the figure, there were other data values with roughly same $z$-values -- $1753.95$ and $1807.26$ in Table~3 (left), $1760.18$ and $1797.53$ in Table~3 (right) -- which were equidistant from the unscattered photon detection point, and exhibiting the highest frequencies. This could be considered as a small drawback in our present data analysis algorithm. \par

\section{\large Discussion}
\label{sec:level6}

We have tried to minimize the computational load during our entire data analysis. Interestingly, we achieved our goal of identifying exactly or near the real scattering points when we consistently followed a set of empirical rules. Following are the empirical rules we framed after several trials with many data samples:

(i) The number of bins \textit{N} was to be chosen beyond the initial fluctuating region ($N>50$) in Fig.~\ref{fig9}.\par

(ii) It was necessary to keep the sample size (\textit{n}) above minimum value ($n\geq 6$) for recognizing the data value correctly for the real scattering point for all SSEs.\par

(iii) The proportion of SSEs in the data sample was needed to be high. This requirement could be problematic if some portions of the data set (sequential event-by-event) contained a relatively high number of USEs and MSEs. In those cases, the proportion of SSEs could be increased by appropriately increasing the sample size (\textit{n}).\par

(iv) Under the above requirements, we succeeded in recognizing the data value for the real scattering points only for those SSEs which had low cardinalities (a small number of possibilities).\par

We did observe the highest frequency for the original (real) scattering or closeby points --  termed as ‘local smoothness’ property -- for the case of the ideal time and energy resolution. In the two bar plots of Fig.~\ref{fig10}, a distinctly high value of the frequency (in red) compared to other nearby values, clearly points towards a significant achievement of our effort considering a sizable statistical fluctuation due to the small sample size (\textit{n}). However, for the cases of finite detector resolutions when the blurring was applied at the post-data processing stage, the results were not as good as for the ideal cases. But, the results remained consistent with the local smoothness property. In effect, we proved the validity of our model and still need to refine the data analysis algorithm.

In the present study, we have avoided showing any enhancement in the reconstructed image as the data analysis algorithm could not handle the total PET data altogether. Instead, the suitable small-sized samples of data were analyzed. A work on devising a proper algorithm capable of analyzing total PET data is currently in progress. It is worth to notice that the real PET data contain various other sources of errors, \textit{e.g.}, random coincidence, non-collinearity, and also the error in attenuation map obtained from CT or MR scan leading to uncertain transformed data. Hence, any new analysis algorithm should possess robustness to handle various errors.

\section{\large Conclusion}
\label{sec:level7}

We have applied the basic concepts of single and double scattering probability distribution functions to develop a model based on the Virtual extrapolation technique. The technique has been applied to a simulated list-mode PET data to create a transformed data set in one variable ($z$). We adopted a thorough randomization procedure in generating the original simulated PET data with Monte Carlo techniques.\par

Our data analysis algorithm relied on a unique `collective difference property' of the transformed data. As the data size turned out to be very large, we followed the procedure of random sampling and sampling distribution. Some empirical observations were made while analyzing the data. We succeeded in finding the real (original) or close by scattering point on the circular locus, which enabled us to retrace the line of response for the single scattered event. So far, the present model has been evaluated for a uniform attenuating phantom medium. We have applied the proposed technique to both TOF- and non-TOF-PETs. After a successful attempt with the perfect timing and energy information, we extended our study for the case of finite resolutions of detectors. Interestingly, uncertain time hardly made any difference because of `length compensation'. Finite energy resolution leads to another symmetric circular locus of scattering points, with increased uncertainties. Moreover, we faced an issue of obtaining another scattering point apart from the real one because of their similar $z$-values. \par

In PET imaging, the addition of single-scattered coincident data is an attractive option to improve image quality because a significant share of PET data is scattered. In that context, we introduced a new approach and proved its success in an initial developmental stage.  For any practical application -- to address the issues arising due to finite energy resolution and mirror symmetry -- we need to extend the present technique for a non-uniform phantom medium.\\

\section*{Acknowledgements}
Authors wish to gratefully acknowledge the Center for Development of Advanced Computing, Pune, India, for providing the supercomputing facilities for data analysis. This research did not receive any specific grant from funding agencies in the public, commercial, or not-for-profit sectors.

\section*{Conflict of interest}
There are no conflicts of interest to declare.
\section*{Ethical approval}
This work does not contain any human and animal study.
\\
\appendix

\setcounter{section}{1}

\appendix

\section{\bf{Concept of compensation length}}
\label{appen1}

The time-of-flight (TOF) information plays a crucial role in assigning the correct position of the annihilation point in PET. However, the TOF information was not needed to perform Virtual extrapolation. We have tried to explain TOF-redundancy property of our model in three ways -- qualitative, simulation based proof, and finally through the results of the analyzed data.\par

TOF redundancy property can be understood from the underlying principle of double coincident scattering in Virtual extrapolation. In the density function $W(z)$ (Eq.~\ref{equ2}), increasing (or decreasing) the distance $l$ in one exponent results in decreasing (or increasing) the distance for the other exponent by an equal amount, causing the net effect to be zero. We termed it `concept of compensation length'. We changed the position of the annihilation point $O$ to any arbitrary point along the line of response $AB$ (Fig.~\ref{fig6}). We then performed Virtual extrapolation, fitted the transformed data with $W(z)$ and found an excellent fit.\par

The implication of the concept of compensation length in our proposed model is profound. Since the TOF information is not needed, the same model can be utilized
for non-TOF-PET systems also.

\section{\bf{Derivation of mean and variance formula for sample statistic $\Delta$}}
\label{appen2}

The mean of sample statistic, $\Delta$ (defined in Sec.~\ref{sec:level4}), is given by,

\begin{equation}
\begin{aligned}
f(n)&=E[\Delta]=E\left [ \sum_{i}\left ( \frac{f_{i}}{n}-w_{i} \right )^{2} \right ],\\
&=\frac{1}{n^{2}}\sum_{i}E[f_{i}^{2}]-\frac{2}{n}\sum_{i}w_{i}E[f_{i}]+\sum_{i}w_{i}^{2}\\
&=\frac{1}{n}\sum_{i}^{N}(w_{i}-w_{i}^{2}),
\end{aligned}
\label{equA1}
\end{equation}

where we used standard formulas for binomial distribution $E[f_{i}]=nw_{i}$ and $E[f_{i}^{2}]=[n(n-1)w_{i}^{2}+nw_{i}]$.
It can be proven that $f_i$ random variables individually obeys binomial distribution, and a set of all random variables $\left \{ f_{1},f_{2},........,f_{N} \right \}$ obey multinomial distribution.\par

The variance of sample statistic, $\Delta$, is given by,

$$V(n)=Var[\Delta]=Var\left [ \sum_{i}\left ( \frac{f_{i}}{n}-w_{i} \right )^{2} \right ],$$

$$V(n)=\sum_{i,j}E\left [ \left ( \frac{f_{i}}{n}-w_{i} \right )^{2}\left ( \frac{f_{j}}{n}-w_{j} \right )^{2} \right ]-\left [ f(n) \right ]^{2}.$$

Simplifying further,

\begin{align*}
V(n)=   &\frac{1}{n^{4}}\sum_{ij}E[f_{i}^{2}f_{j}^{2}]-\frac{2}{n^{3}}\sum_{ij}w_{j}E[f_{i}^{2}f_{j}]\\
        &+\frac{1}{n^{2}}\sum_{ij}w_{j}^{2}E[f_{i}^{2}]-\frac{2}{n^{3}}\sum_{ij}w_{i}E[f_{i}f_{j}^{2}]\\
        &+\frac{4}{n^{2}}w_{i}w_{j}E[f_{i}f_{j}]-\frac{2}{n}\sum_{ij}w_{i}w_{j}^{2}E[f_{i}]\\
        &+\frac{1}{n^{2}}\sum_{ij}w_{i}^{2}w_{j}^{2}-\frac{1}{n^{2}}\left [ \sum_{i}(w_{i}-w_{i}^{2}) \right]^{2}.
\end{align*}

Since,

$$\frac{1}{n^{2}}\left [ \sum_{i}(w_{i}-w_{i}^{2}) \right]^{2}=\frac{1}{n^{2}}\sum_{ij}(w_{i}-w_{i}^{2})(w_{j}-w_{j}^{2}),$$

$$E[f_{i}f_{j}]=n(n-1)w_{i}w_{j},$$

$$E[f_{i}^{2}f_{j}]=n(n-1)(n-2)w_{i}^{2}w_{j}+n(n-1)w_{i}w_{j},$$

$$E[f_{i}f_{j}^{2}]=n(n-1)(n-2)w_{i}w_{j}^{2}+n(n-1)w_{i}w_{j},$$

\begin{align*}
  E[f_{i}^{2}f_{j}^{2}]&=n(n-1)(n-2)(n-3)w_{i}^{2}w_{j}^{2}\\
  &+n(n-1)(n-2)w_{i}^{2}w_{j}\\
  &+n(n-1)(n-2)w_{i}w_{j}^{2}\\
  &+n(n-1)w_{i}w_{j}.
\end{align*}

After substituting above values in the expression for $V(n)$, we get the formula for the variance of $\Delta$,

\begin{equation}
\begin{aligned}
V(n)&=Var[\Delta]=\left ( \frac{2}{n^{2}}-\frac{6}{n^3} \right )\sum_{i=1,j=1}^{N}w_{i}^{2}w_{j}^{2}\\
    &+\left ( \frac{4}{n^{3}}\right )\sum_{i=1,j=1}^{N}w_{i}^{2}w_{j}- \left (\frac{1}{n^{3}} \right )\sum_{i=1,j=1}^{N}w_{i}w_{j}.
\end{aligned}
\label{equA2}
\end{equation}



\bibliographystyle{model3-num-names}
\bibliography{mybibfile1.bib}

\end{document}